\documentclass{aa} 
\usepackage{natbib} 
\usepackage{graphics}
\usepackage{amssymb} 

\begin{document}

\title{The impact of shocks on the chemistry of molecular clouds}
\subtitle{High resolution images of chemical differentiation along the NGC1333-IRAS2A outflow}

\author{J.K. J{\o}rgensen\inst{1} \and M.R. Hogerheijde\inst{2} \and
G.A. Blake\inst{3} \and E.F. van Dishoeck\inst{1} \and
L.G. Mundy\inst{4} \and F.L. Sch\"{o}ier\inst{1,5}}

\institute{Leiden Observatory, P.O. Box 9513, NL-2300 RA Leiden, The
Netherlands \and Steward Observatory, The University of Arizona, 933
N. Cherry Avenue, Tucson, AZ 85721-0065, USA \and Division of
Geological and Planetary Sciences, California Institute of Technology,
MS 150-21, Pasadena, CA 91125, USA \and Department of Astronomy,
University of Maryland, College Park, MD 20742, USA \and Stockholm
Observatory, AlbaNova, SE-106 91 Stockholm, Sweden}

\offprints{Jes K.\,J{\o}rgensen} \mail{joergensen@strw.leidenuniv.nl}
\date{Received <date> / Accepted <date>}

\abstract{This paper presents a detailed study of the chemistry in the
outflow associated with the low-mass protostar NGC~1333-IRAS2A down to
3\arcsec\ (650 AU) scales. Millimeter-wavelength aperture-synthesis
observations from the Owens Valley and
Berkeley-Illinois-Maryland-Association interferometers and
(sub)millimeter single-dish observations from the Onsala Space
Observatory 20~m telescope and Caltech Submillimeter Observatory are
presented. The interaction of the highly collimated protostellar
outflow with a molecular condensation $\sim$15000~AU from the central
protostar is clearly traced by molecular species such as HCN, SiO, SO,
CS, and CH$_3$OH. Especially SiO traces a narrow high velocity
component at the interface between the outflow and the molecular
condensation.  Multi-transition single-dish observations are used to
distinguish the chemistry of the shock from that of the molecular
condensation and to address the physical conditions
therein. Statistical equilibrium calculations reveal temperatures of
20 and 70~K for the quiescent and shocked components, respectively,
and densities near $10^6$~${\rm cm}^{-3}$. The line-profiles of low-
and high-excitation lines are remarkably similar, indicating that the
physical properties are quite homogeneous within each
component. Significant abundance enhancements of two to four orders of
magnitude are found in the shocked region for molecules such as
CH$_3$OH, SiO and the sulfur-bearing molecules. HCO$^+$ is seen only
in the aftermath of the shock consistent with models where it is
destroyed through release of H$_2$O from grain mantles in the
shock. N$_2$H$^+$ shows narrow lines, not affected by the outflow but
rather probing the ambient cloud. The overall molecular inventory is
compared to other outflow regions and protostellar
environments. Differences in abundances of HCN, H$_2$CO and CS are
seen between different outflow regions and are suggested to be related
to differences in the atomic carbon abundance. Compared to the warm
inner parts of protostellar envelopes, higher abundances of in
particular CH$_{3}$OH and SiO are found in the outflows, which may be
related to density differences between the regions.

} \maketitle

\section{Introduction}\label{introduction}
One of the manifestations of a newly formed low-mass protostar is the
presence of a highly collimated and energetic outflow or jet. A
natural consequence of the propagation of such high velocity outflows
through the protostellar envelope and the ambient molecular medium are
shockfronts \citep{reipurth99}. Shocks both heat and compress the gas
and also trigger chemical reactions in the gas-phase, leading to a
different chemistry than observed otherwise. Shock processing of dust
grains may lead to the injection of atoms and molecules back into the
gas, which further distinguishes the chemistry in the shocked region
from that of a quiescent protostellar environment
\citep[e.g.,][]{tielens99}. This paper presents a study of the physics
and chemistry of the outflow associated with a well-known class 0
young stellar object \object{NGC~1333-IRAS2} using high-resolution
millimeter wavelength interferometer and multi-transition single-dish
observations. The \object{NGC~1333-IRAS2} outflow provides a unique
opportunity to study the effects of outflows on ambient molecular
clouds, as the main shock is well separated from the central protostar
and shows a relatively simple morphology. The combination of
single-dish and interferometry observations makes it possible to
discuss the physical and chemical properties of the outflowing gas and
to address the spatial differentiation of the chemistry in the outflow
region resolved by the interferometer observations.

Studies of molecular abundances in regions of high outflow activity
provide insight into the dependence of the chemical reaction networks
on temperature and density. Furthermore it is important to recognize
the effect of outflow-triggered chemistry in the inner protostellar
envelope, to disentangle it from emission from a circumstellar disk or
to address the effect of passive heating by the central protostar. In
the central part of the protostellar envelope, thermal evaporation of
dust grain mantles can lead to a distinct chemistry as is seen in the
case of low-mass protostars \citep[e.g
IRAS~16293-2422,][]{ceccarelli00a,ceccarelli00b,schoeier02,cazaux03}.

\begin{figure*}
\resizebox{\hsize}{!}{\rotatebox{90}{\includegraphics{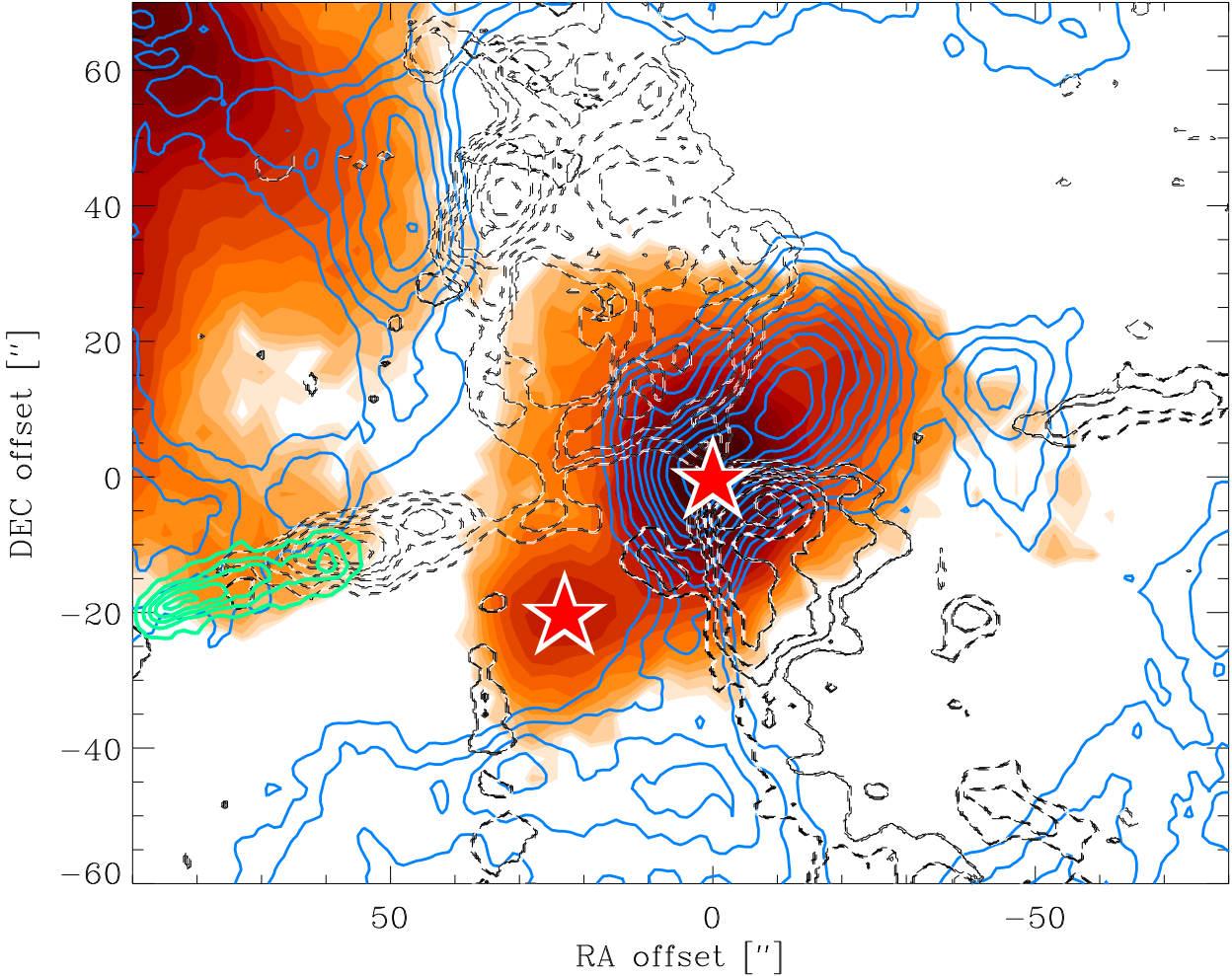}}}
\caption{Overview of the \object{IRAS2A} outflow region. The colored
image shows the SCUBA 850~$\mu$m emission tracing the cold
dust. N$_2$H$^+$ BIMA line observations and SiO OVRO line observations
(this paper) are indicated by blue and green contours,
respectively. CO 2-1 emission from the \cite{engargiola99} is
indicated by the black/white dashed contours. The red stars indicate
the position of IRAS2A and IRAS2B from
\cite{n1333i2art}.}\label{scuba_overview}
\end{figure*}

\object{NGC~1333-IRAS2} (also known as \object{IRAS~03258+3104};
hereafter simply \object{IRAS2}) is located in the \object{NGC~1333}
molecular cloud, harboring several class 0 and I objects, first
identified through IRAS maps by \cite{jennings87}. Continuum
observations reveal that \object{IRAS2} is a binary source with two
components, \object{IRAS2A} and 2B, separated by 6500~AU (30\arcsec)
\citep{sandell94knee,blake96,looney00}. \object{IRAS2A} is responsible
for a highly collimated east-west outflow giving rise to a strong
shock $\sim$15000~AU from the central continuum source
(Fig.~\ref{scuba_overview}). A strong CO outflow in the north-south
direction has also been observed \citep{liseau88,engargiola99}, which
originates within a few arcseconds from \object{IRAS2A}
\citep{n1333i2art}.

\cite{langer96} mapped the entire \object{NGC~1333} region in CS and
identified two peaks in CS emission toward the \object{IRAS2}
outflow. They suggested that these are associated with red-shifted
(eastern) and blue-shifted (western) bow shock components of the
outflow. \cite{sandell94knee} reported bright CH$_{3}$OH emission toward
the red-shifted (eastern) outflow lobe and correspondingly a high
CH$_{3}$OH abundance, possibly enhanced by the shock. The more detailed
structure of the CH$_{3}$OH emission from the \object{IRAS2A} east-west
outflow was discussed by \cite{bachiller98}, who mapped the outflow
positions at $\approx$~3\arcsec\ using the IRAM interferometer and
30~m single-dish telescope. \citeauthor{bachiller98} derived the
physical conditions in the shock interaction zone from LVG
calculations and obtained a density of $\sim 10^6$~cm$^{-3}$ and
temperature of $\sim 100$~K.

\citeauthor{bachiller98} also found that the observed methanol
emission translates to a large enhancement of CH$_3$OH by a factor
$\sim 300$ in the \object{IRAS2A} outflow. CH$_{3}$OH is thought to be
released directly from the dust grain mantles and is often seen to be
associated with protostellar outflows
\citep[e.g.,][]{bachiller95}. Other often-used tracers of shocks
associated with protostellar outflows are Si-bearing species, in
particular SiO \citep[e.g.,][]{martinpintado92,codella99,garay02}. High
abundances of these species may mark a clear distinction of shocked
gas from unprocessed gas in the envelopes around low-mass protostars
\citep{bachiller97,garay98,bachiller01}.

In this paper we present a study of the detailed chemistry of the
shock associated with the \object{IRAS2A} outflow based on
observations of a wide range of molecular lines at $\sim 3-6$\arcsec\
resolution from the Owens Valley Radio Observatory (OVRO) and Berkeley
Maryland Illinois Association (BIMA) millimeter interferometers,
together with millimeter and submillimeter single dish observations
from the Onsala 20~m telescope (OSO) and the Caltech Submillimeter
Observatory 10.4~m telescope (CSO). Parts of the OVRO observations
have previously been presented by
\cite{blake96}. Sect.~\ref{observations} describes the observations
and reductions. The maps from the interferometry observations are
presented and discussed in Sect.~\ref{interferometry}, while the
single dish observations are treated in Sect.~\ref{singledish}. The
physical and chemical properties of the shock region are analyzed
using statistical equilibrium calculations as described in
Sect.~\ref{analysis} and molecular abundances are
derived. Sect.~\ref{discussion} discusses the inferred chemistry and
compares it to other well-studied outflow regions, to other types of
star-forming environments and to available models for the chemistry in
outflow regions. The main findings are summarized in
Sect.~\ref{conclusion}. A companion paper \citep{n1333i2art} presents
details of a millimeter-wavelength interferometer study of the
environment surrounding the central protostellar system.

\section{Overview of observations\label{observations}}
The position of the shock in the eastern lobe of the outflow
associated with \object{IRAS2A} ($\alpha(2000)=03^{\rm h}29^{\rm
m}00\fs0$; $\delta(2000)=31^\circ14\arcmin19\farcs0$) was observed
with the Millimeter Array of the Owens Valley Radio Observatory
(OVRO)\footnote{The Owens Valley Millimeter Array is operated by the
California Institute of Technology under funding from the US National
Science Foundation (grant no. AST-9981546).} between October 5, 1994
and January 1, 1995 in the six-antenna L- and H-configurations. Tracks
were obtained in two frequency settings at 86 and 97~GHz, and each
track observed in alternatingly two fields: the bow shock at the end
of the eastern outflow discussed in this paper and the position of the
central protostellar source \citep{n1333i2art}. The observed tracks
cover projected baselines of 3.1--70~k$\lambda$ at 86~GHz and the lines
observed are listed in Table~\ref{obs_overview}. The lines were
recorded in spectral bands with widths of 32~MHz ($\sim
100$~km~s$^{-1}$). H$^{13}$CO$^+$ $1-0$ and CS $2-1$ were observed in 128
spectral channels, the remaining line setups included 64 spectral
channels. The complex gain variations were calibrated by observing the
nearby quasars 0234+285 and 3C84 approximately every 20
minutes. Fluxes were calibrated by observations of Uranus and
Neptune. The rms noise levels are 0.05~Jy~beam$^{-1}$ in the 250 kHz
channels with a synthesized beam size of $3.2\times
2.8$\arcsec. Calibration and flagging of visibilities with clearly
deviating amplitudes and/or phases was performed with the MMA
reduction package \citep{scoville93}.

The millimeter interferometer of the Berkeley-Illinois-Maryland
Association (BIMA)\footnote{The BIMA array is operated by the
Universities of California (Berkeley), Illinois, and Maryland, with
support from the National Science Foundation (grants AST-9981308,
AST-9981363 and AST-9981289).} observed the \object{IRAS2A} outflow
position between March 4 and April 15, 2003. The array B- and
C-configurations provided projected baselines of
2.7--71~k$\lambda$. The lines of HCO$^+$ 1--0, HCN 1--0, N$_2$H$^+$
1--0, and C$^{34}$S 2--1 were recorded in 256-channel spectral bands
with a total width of 6.25~MHz ($\sim 20$~km~s$^{-1}$). The complex
gain of the interferometer was calibrated by observing the bright
quasars 3C84 (4.2~Jy) and 0237+288 (2.3~Jy) approximately every 20
minutes. The absolute flux scale was bootstrapped from observations of
Uranus. The rms noise levels are 0.2~Jy~beam$^{-1}$ in the 24~kHz
channels, with a synthesized beam size of $6.1''\times 5.0''$ FWHM
($7.6''\times 6.8''$ for the C$^{34}$S and N$_2$H$^+$ observations). The data
were calibrated with routines from the MIRIAD software package
\citep{sault95}.

In addition to the interferometry data, a number of molecular lines
were observed toward the position of the red-shifted shock using the
Caltech Submillimeter Observatory 10.4~m (CSO)\footnote{The Caltech
Submillimeter Observatory 10.4~m is operated by Caltech under a
contract from the National Science Foundation (grant
no. AST-9980846).} and Onsala Space Observatory 20~m
(OSO)\footnote{The Onsala 20~m telescope is operated by the Swedish
National Facility for Radio Astronomy, Onsala Space Observatory at
Chalmers University of Technology.}  telescopes. The pointing was
checked regularly and found to be accurate to a few arcseconds. The
typical beam sizes are 45\arcsec--33\arcsec\ for the OSO 20~m
(86--115~GHz) and 26\arcsec--20\arcsec\ for the CSO (217--356~GHz)
observations. The data were calibrated using the standard chopper
wheel method. The spectra were reduced in a standard way by
subtracting baselines and by dividing by the main-beam efficiencies
$\eta_{\rm mb}$ as given on the web pages for the two
telescopes. $\eta_{\rm mb}$ ranges from 0.6 to 0.43 for frequencies of
86 to 115~GHz for the OSO 20~m and 0.67 to 0.62 for frequencies of 217
to 356~GHz for the CSO. An overview of all the observed lines
(single-dish and interferometer) is given in Table~\ref{obs_overview}.

\begin{table}
\caption{Overview of the observations of the \object{IRAS2A} outflow
position treated in this paper.}\label{obs_overview}
\begin{tabular}{lllll}\hline\hline
Line & & Rest freq.& Observed with \\ \hline
CO          & $2-1$         & 230.5380             & CSO         \\
            & $3-2$         & 345.7960             & CSO         \\[0.5ex]
C$^{18}$O       & $3-2$         & 329.3305             & CSO         \\[0.5ex]
CH$_3$OH    & $2_1-1_1$     & \phantom{1}97.5828   & OVRO        \\
            & $7_2-6_1$     & 338.7222             & CSO         \\[0.5ex]
CS          & $2-1$         & \phantom{1}97.9810   & OSO, OVRO   \\
            & $5-4$         & 244.9356             & CSO         \\
            & $7-6$         & 342.8830             & CSO         \\[0.5ex]
C$^{34}$S   & $2-1$         & \phantom{1}96.4129   & BIMA        \\[0.5ex] 
HCN         & $1-0$$^{a}$ & \phantom{1}88.6318   & OSO, BIMA   \\
            & $4-3$         & 354.5055             & CSO         \\[0.5ex]
HCO$^+$     & $1-0$         & \phantom{1}89.1885   & OSO, BIMA   \\
            & $4-3$         & 356.7343             & CSO         \\[0.5ex]
H$^{13}$CO$^+$     & $1-0$         & \phantom{1}86.7543   & OSO, OVRO   \\[0.5ex]
H$_2$CO     & $5_{1,5}-4_{1,4}$ & 351.7686         & CSO         \\
            & $5_{0,5}-4_{0,4}$ & 362.7359         & CSO         \\[0.5ex]
N$_2$H$^+$  & $1-0$$^{a}$ & \phantom{1}93.1737   & OSO, BIMA   \\[0.5ex]
SiO         & $2-1$         & \phantom{1}86.8470   & OSO, OVRO   \\
            & $5-4$         & 217.1049             & CSO         \\
            & $8-7$         & 347.3306             & CSO         \\[0.5ex]
SO          & $2_2-1_1$     & \phantom{1}86.0940   & OVRO        \\
            & $2_3-1_2$     & \phantom{1}99.2999   & OSO         \\
            & $8_9-7_8$     & 346.5285             & CSO         \\[0.5ex]
SO$_2$      & $7_{3,5}-8_{2,6}$ & 97.7024      & OVRO        \\\hline
\end{tabular}

$^{a}$Hyperfine splitting - multiple lines observed in one
setting. The coordinates for the single-dish observations are
$\alpha(2000)=03^{\rm h}29^{\rm m}01\fs0$,
$\delta(2000)=31^\circ14\arcmin20\arcsec$, i.e., corresponding to
offset of (79\arcsec,-17\arcsec) in the maps presented in this paper.
\end{table}

\section{Data}\label{data}
\subsection{Interferometry}\label{interferometry}
\begin{figure*}
\centering
\resizebox{\hsize}{!}{\includegraphics{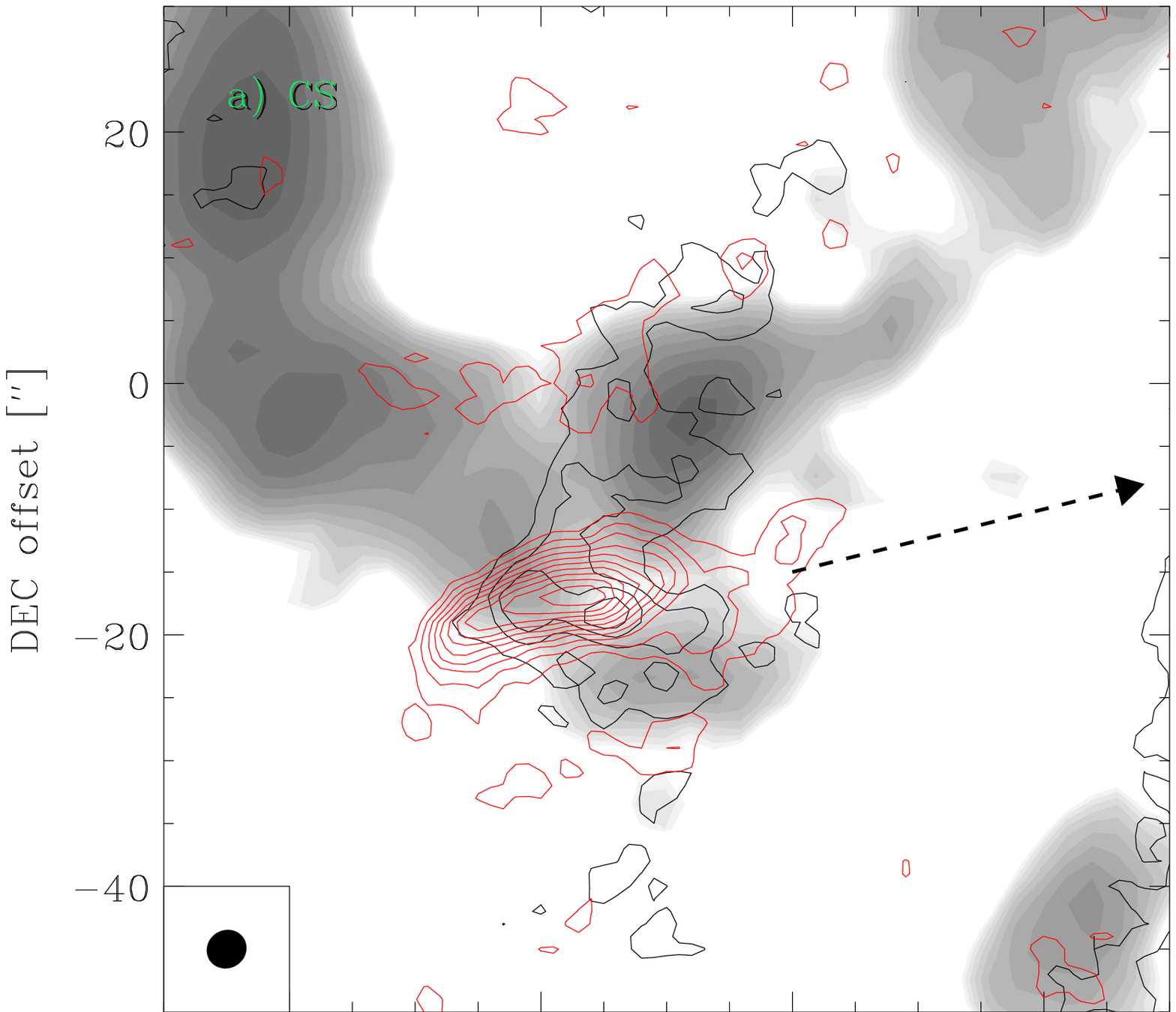}\includegraphics{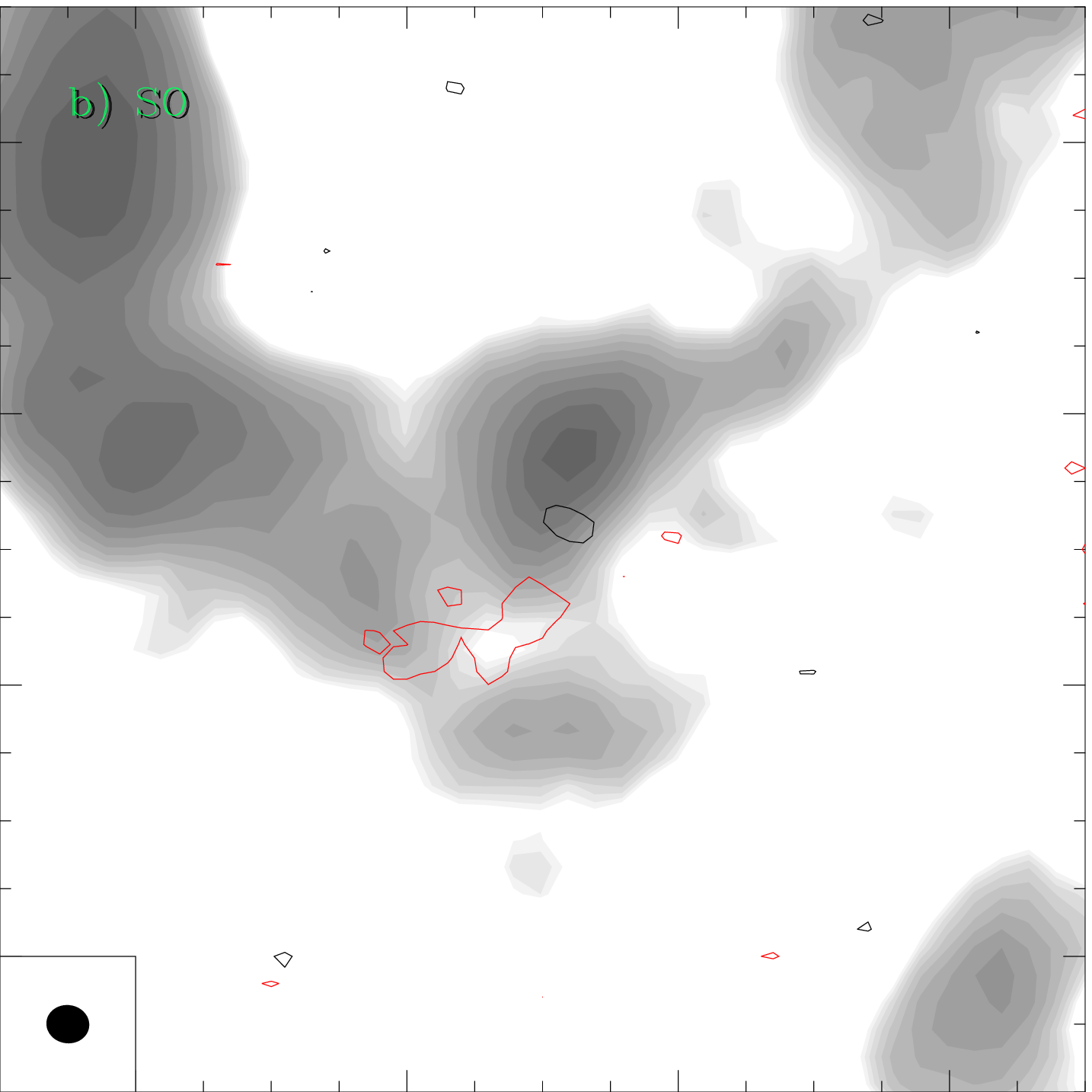}}
\resizebox{\hsize}{!}{\includegraphics{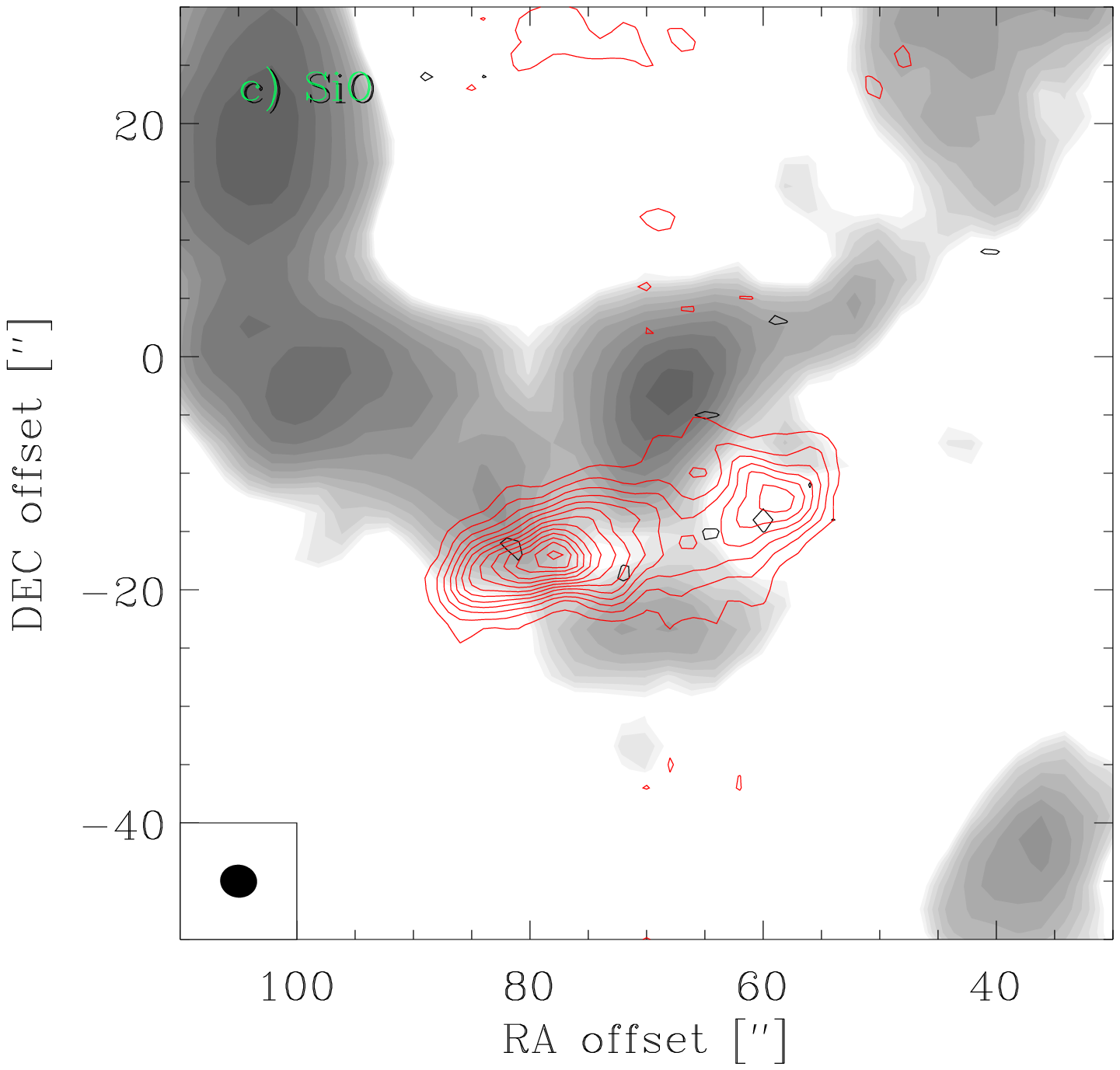}\includegraphics{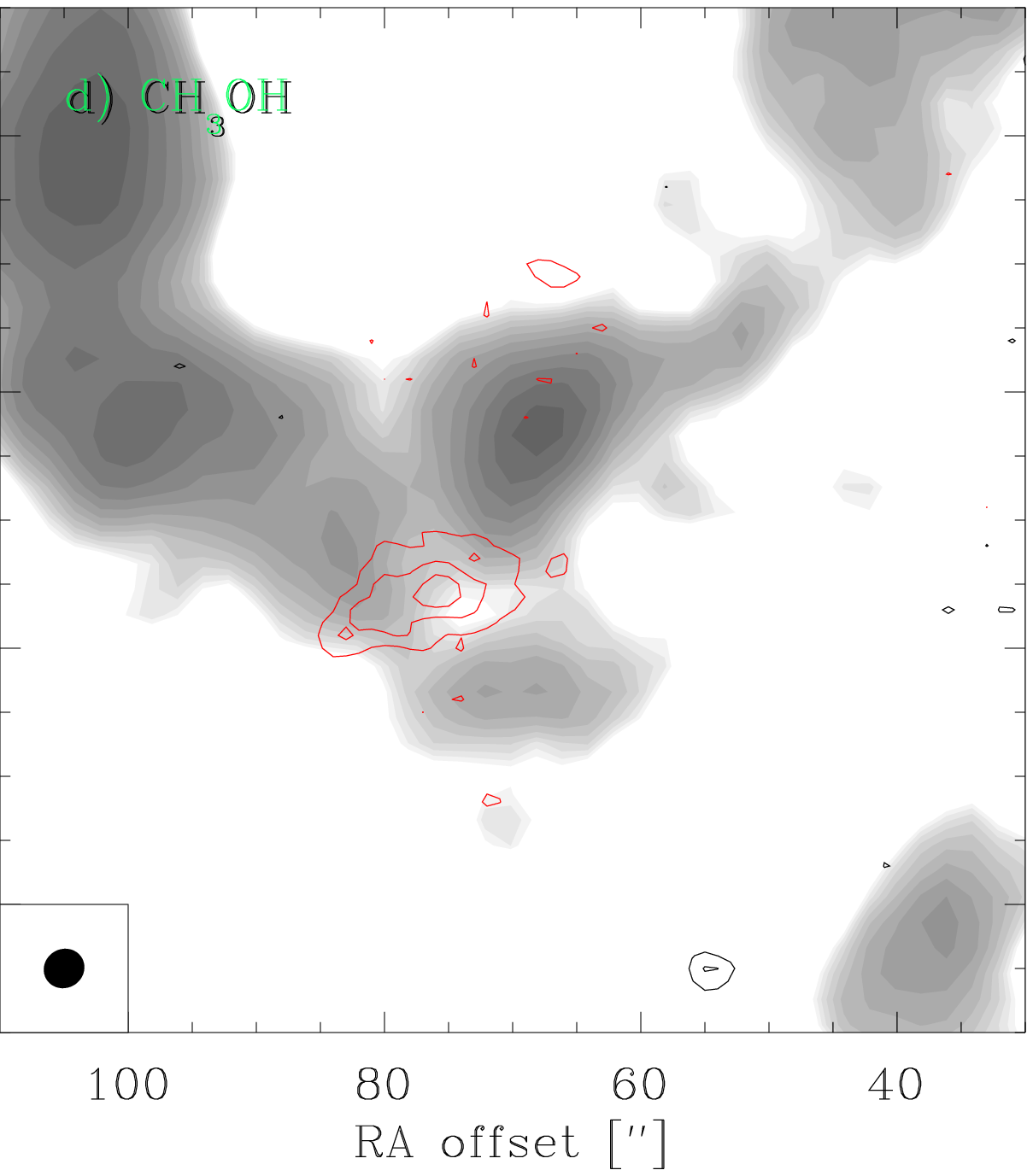}}
\caption{Moment maps for the lines observed at OVRO: a) CS, b) SO, c)
SiO and d) CH$_{3}$OH. The black lines indicate low velocity emission
integrated over velocities of 5-9~km~s$^{-1}$ whereas the red lines indicate
high velocity emission integrated over 9-16~km~s$^{-1}$ for CH$_{3}$OH and SO
and 9-25~km~s$^{-1}$ for CS and SiO. The contours are given in steps of
3$\sigma$ and are overlaid on grey-scale images of the low velocity
N$_2$H$^+$ emission. The x- and y-axis offsets are given relative to
the position of the \object{IRAS2A} source. In the upper left panel
the arrow indicates the direction back to the central
protostar.}\label{ovro_moments}
\end{figure*}

\begin{figure*}
\resizebox{\hsize}{!}{\includegraphics{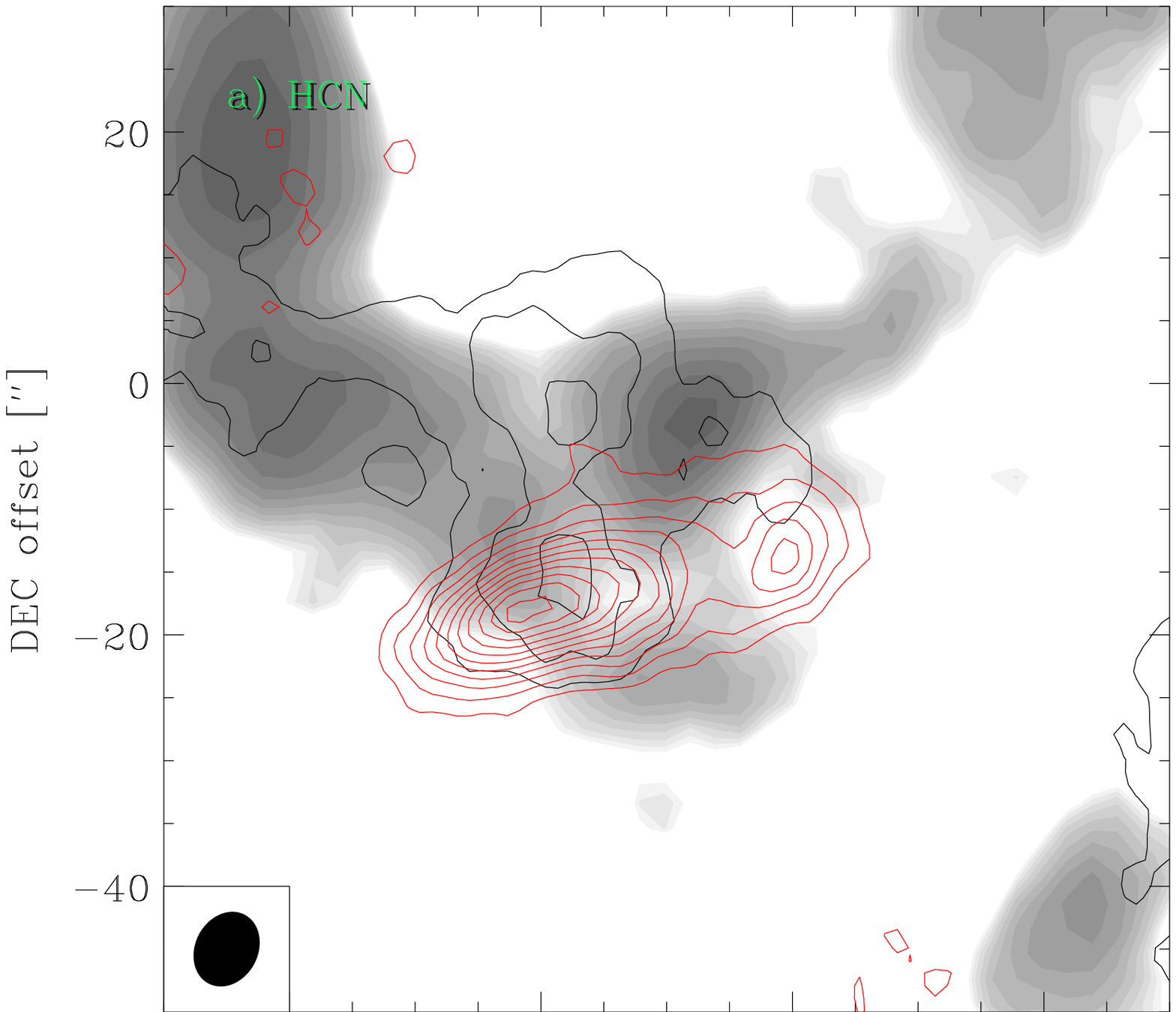}\includegraphics{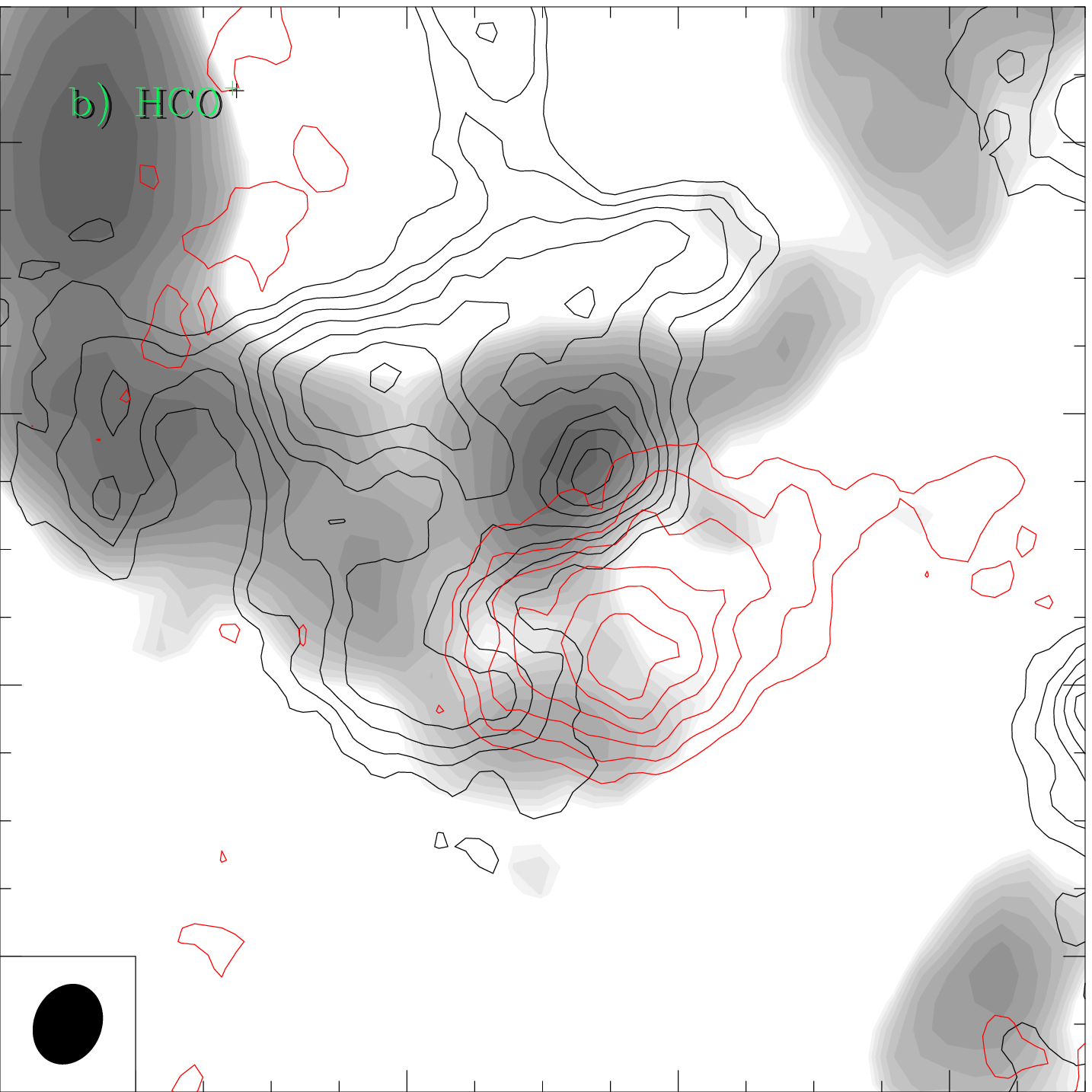}}
\resizebox{\hsize}{!}{\includegraphics{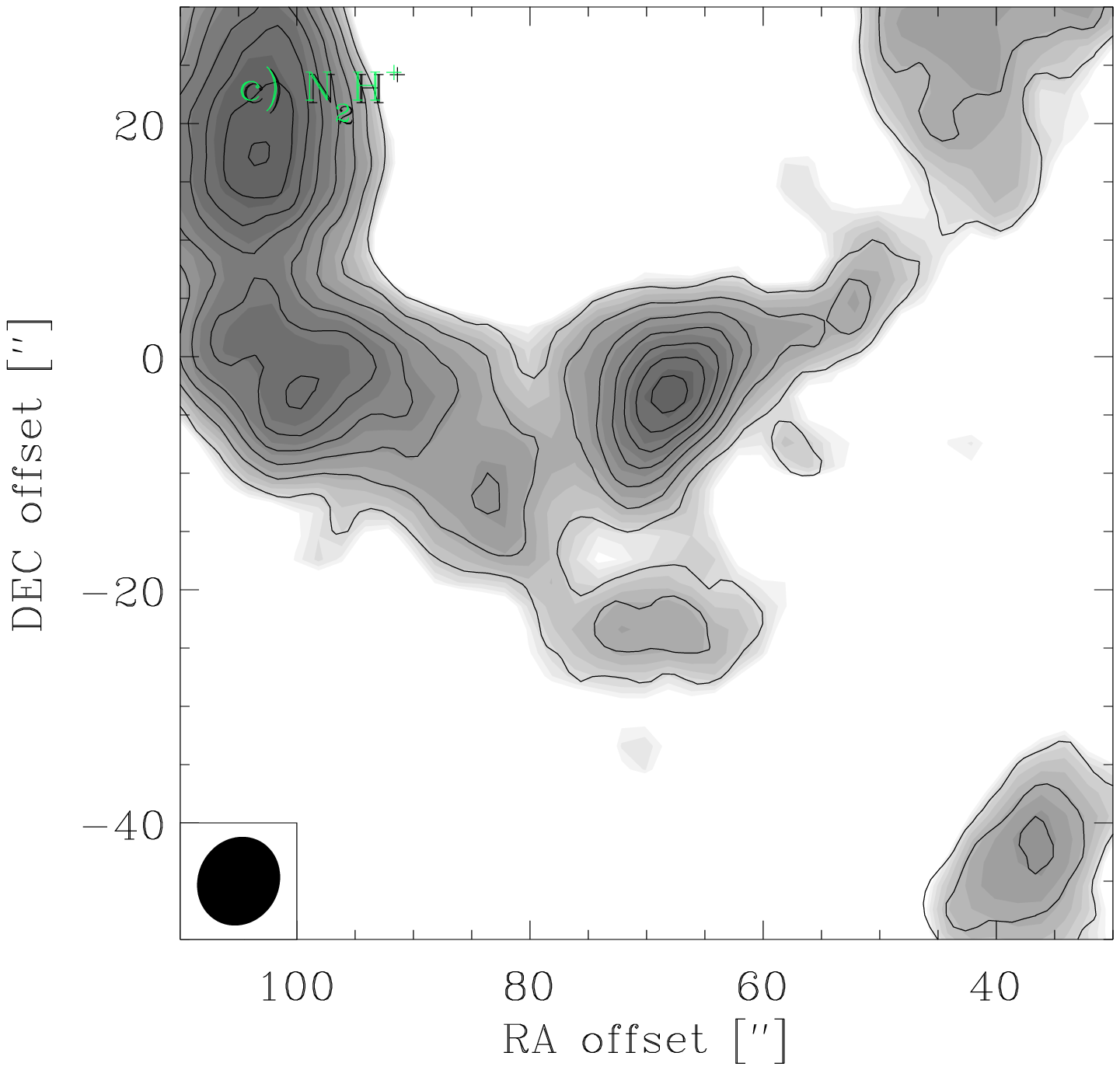}\includegraphics{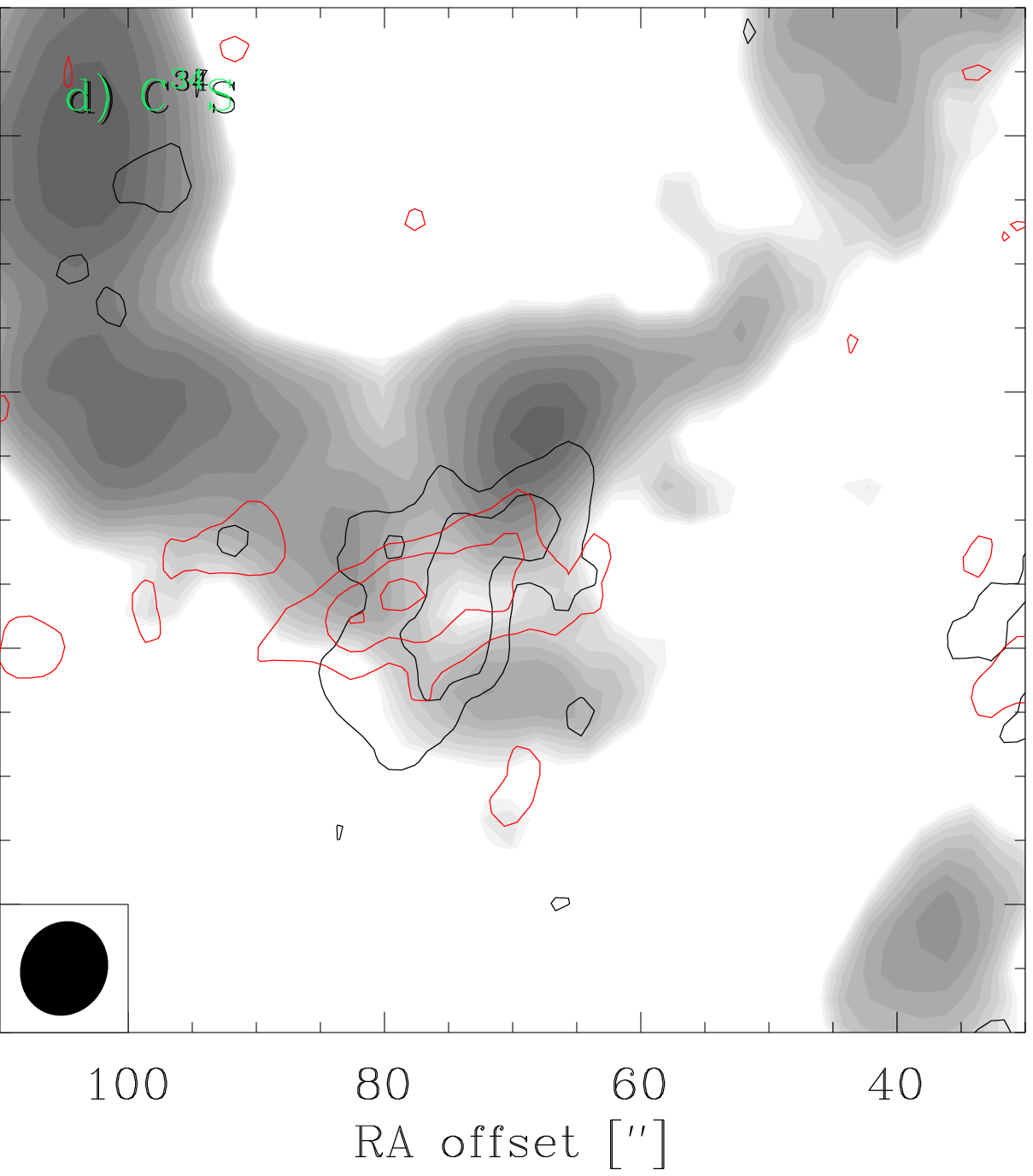}}
\caption{Moment maps for the lines observed at BIMA: a) HCN, b) HCO$^+$,
c) N$_2$H$^+$ and d) C$^{34}$S. As in Fig.~\ref{ovro_moments} the black
and red lines indicate low and high velocity material, respectively
(the low velocity emission is integrated over 5-9~km~s$^{-1}$ and the high
velocity emission over 9-16~km~s$^{-1}$). The contours ascend in steps of
3$\sigma$ and are overlaid on grey-scale images of the low velocity
N$_2$H$^+$ emission.}\label{bima_moments}
\end{figure*}

Fig.~\ref{ovro_moments}-\ref{bima_moments} show moment maps for the
lines observed at OVRO (CS, SO, SiO and CH$_{3}$OH) and BIMA (i.e.,
HCO$^+$, HCN, N$_2$H$^+$ and C$^{34}$S). In all maps the coordinates are
given as offsets relative to the position of the central protostar,
\object{IRAS2A}: $\alpha(2000)=03^{\rm h}28^{\rm m}55\fs7$,
$\delta(2000)=31^\circ14\arcmin37\arcsec$. Emission of SO$_2$ and
H$^{13}$CO$^+$ was not detected in the interferometer maps toward the
outflow position.

Most of the observed lines show indications of material affected by
the outflow with clear line wings spreading out to 10-15~km~s$^{-1}$ from
the systemic velocity of 7~km~s$^{-1}$. One exception is N$_2$H$^+$ which shows
narrow hyperfine components of approximately 1~km~s$^{-1}$ width (FWHM). Of
the observed lines, SiO, CS and HCN show emission stretching furthest
from the systemic velocity (out to $\approx$~20~km~s$^{-1}$) while the
remaining species show somewhat narrower profiles (wings stretching
out to $\approx$~10~km~s$^{-1}$ relative to the systemic velocity), with
HCO$^+$ showing most material closest to the cloud systemic
velocity. SiO is not seen at low velocities as is also the case for
CH$_{3}$OH and SO.

In general the molecules observed at BIMA most clearly trace the
extended low velocity material. This may in part be due to the
different $(u,v)$ coverage of the two arrays, with the BIMA observations
recovering more of the extended emission, but the species observed at
OVRO may also be those that are predominantly enhanced by the outflow
generated shock.

The N$_2$H$^+$ emission traces a ridge of material with a number of
``cores'' stretching from the north-east of the map toward the center
and back again to the north-west. A dominant core is seen close to the
center (offsets of (67\arcsec,-3\arcsec)) which is also picked up by
the HCO$^+$ maps. South of this core a ``$<$''-shaped extension is seen
in both HCO$^+$ and N$_2$H$^+$. The low velocity CS emission picks up only
this feature. A similar component was also seen in the CH$_{3}$OH
emission mapped by \cite{bachiller98} but is not evident in the
CH$_{3}$OH observations presented here, possibly due to lower
sensitivity.

The high-velocity material is generally much less extended than the
low-velocity material. The HCN, SiO and CS trace a narrow component
stretching 30-40\arcsec\ along the outflow propagation and
5-10\arcsec\ in the direction perpendicular to this. The narrow
component points directly to the ``$<$''-shaped feature in the low
velocity emission material. The HCN emission is slightly more extended
than that of the two other species, again likely due to the different
$(u,v)$ coverage of the observations from the two arrays. SO and
CH$_{3}$OH show a slightly weaker structure along the same narrow
component.

The emission of CH$_{3}$OH and SO is located downstream (west) of the
outflow propagation direction compared to, e.g., the peak of SiO. Even
farther downstream around offset (59\arcsec,-13\arcsec), HCN, CS and
SiO show another strong feature where the N$_2$H$^+$ emission ``pinches''
the outflow. The HCO$^+$ wing emission is seen only at this position
and is found to be more extended, filling out the region void of
N$_2$H$^+$. In fact the HCO$^+$ emission can be traced all the way back to
the central protostar as is also the case for CO
(Fig.~\ref{scuba_overview}). It is striking how the HCO$^+$ and, e.g.,
HCN wing emission trace significantly different components, implying a
clear chemical differentiation.

\begin{figure*}
\resizebox{\hsize}{!}{\includegraphics{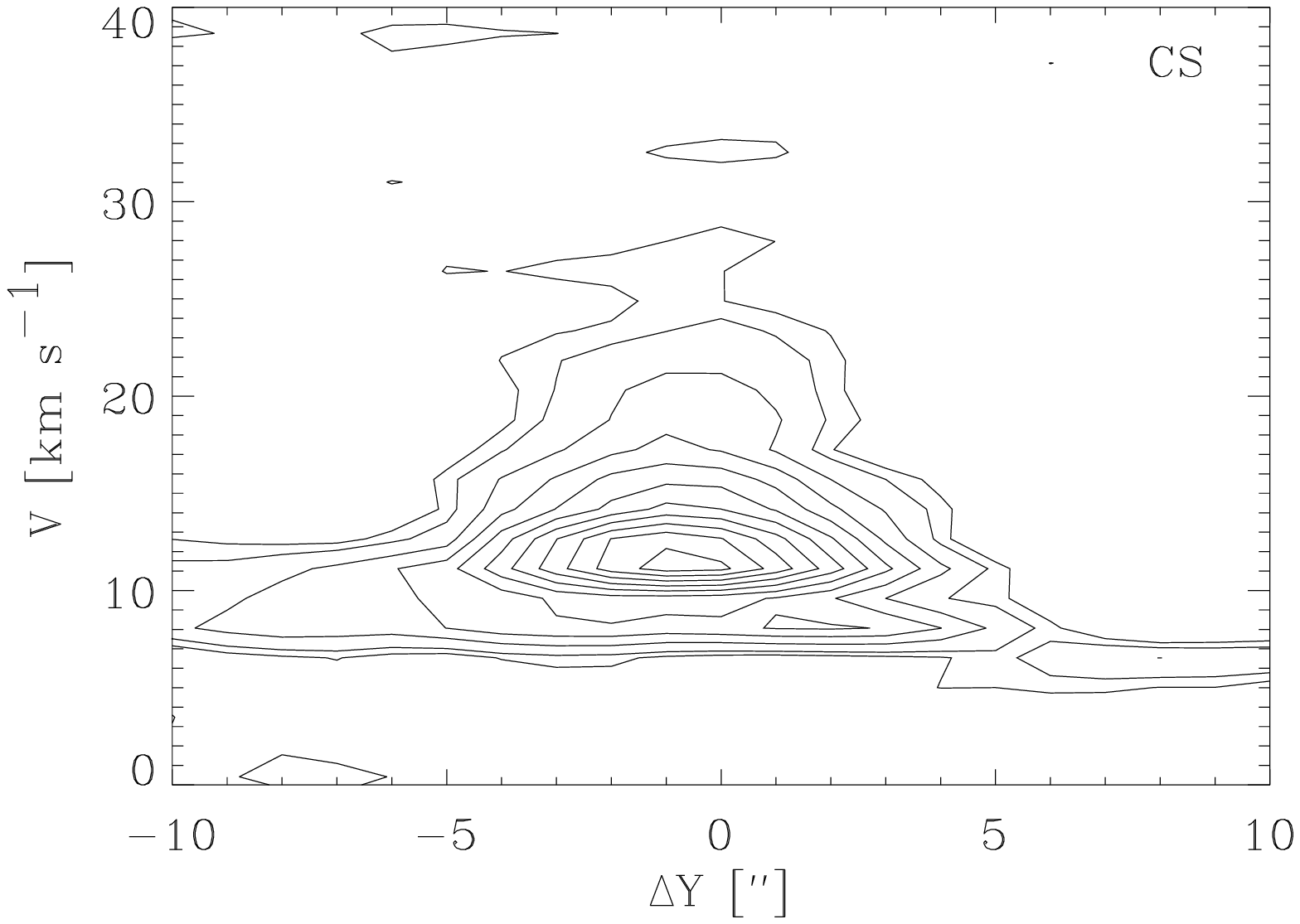}\includegraphics{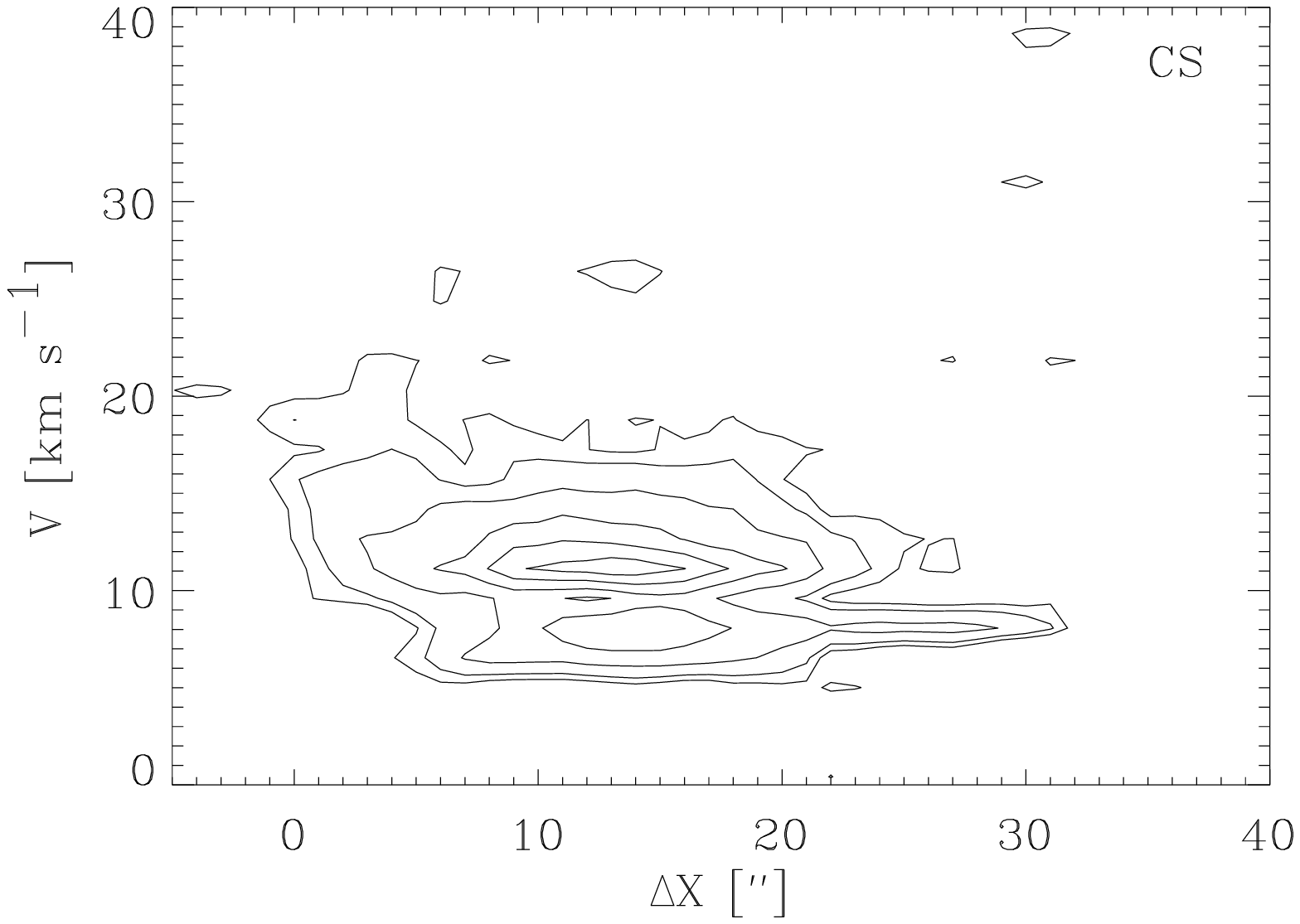}}
\resizebox{\hsize}{!}{\includegraphics{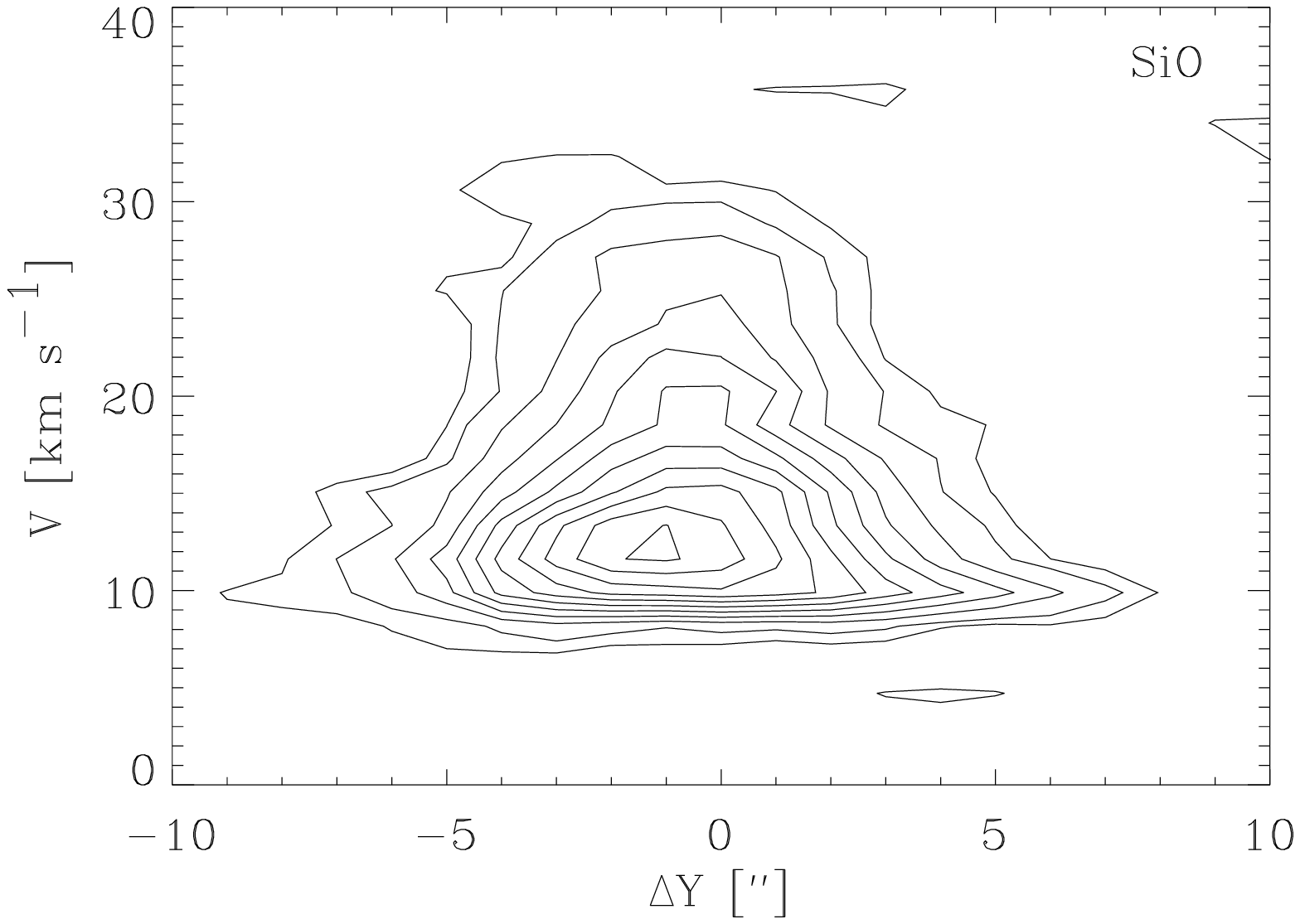}\includegraphics{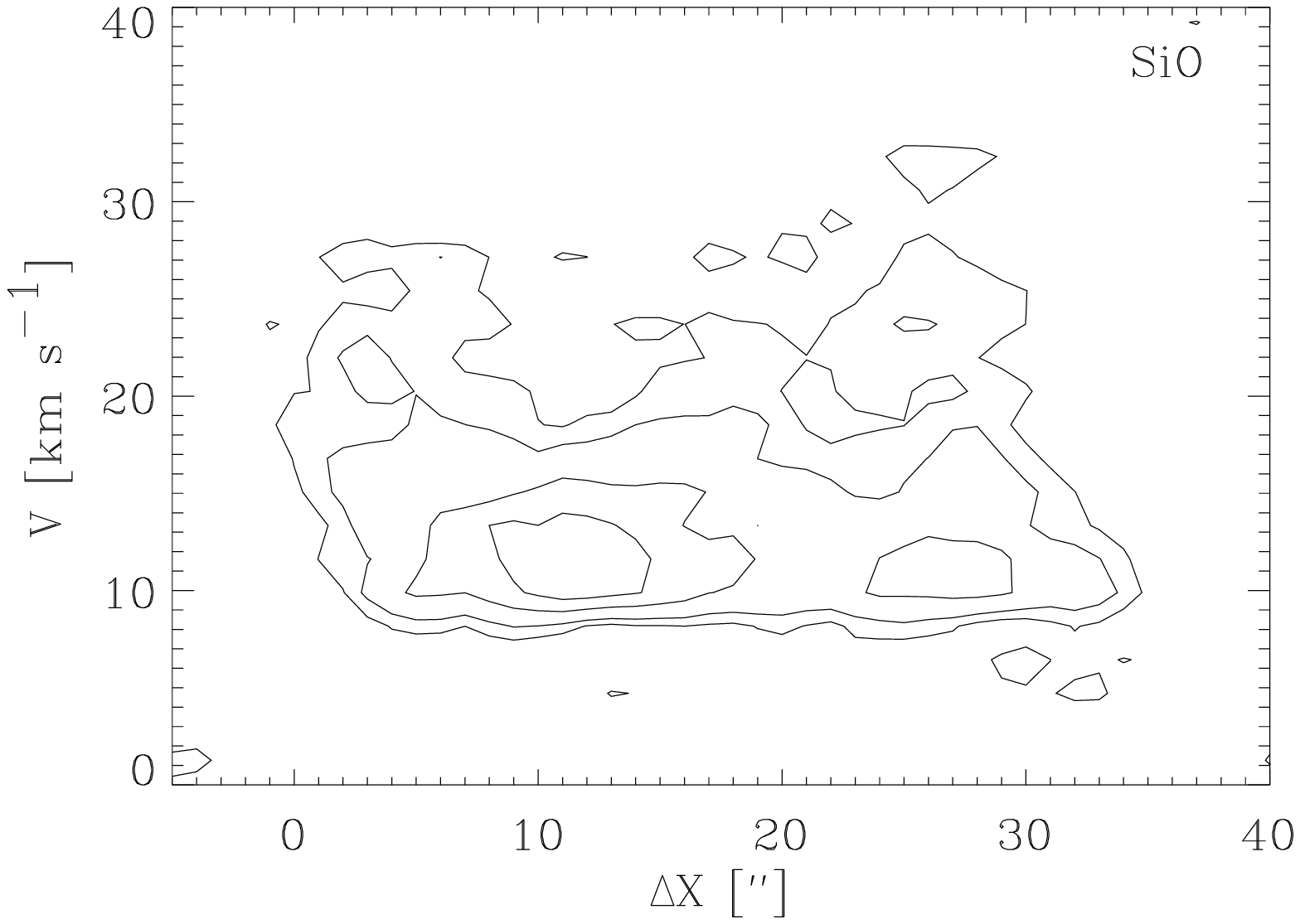}}
\caption{CS (upper) and SiO (lower) position-velocity maps for the
IRAS2A shock. The coordinate frame has been rotated and translated to
an XY-coordinate system with the X-axis along the propagation
direction of the outflow ($\sim$ 19$^\circ$ with the RA axis in
Fig.~\ref{ovro_moments}) and the Y-axis in the perpendicular/north
direction. The (0,0) point for the XY coordinate system has been
chosen to be at (87\arcsec,-23\arcsec) i.e., at the working surface,
or head, of the outflow as judged from the morphology of the high
velocity emission. The contours are given at 2$\sigma$, 4$\sigma$,
8$\sigma$, $\ldots$ and upwards in steps of
4$\sigma$.}\label{pv_diagram}
\end{figure*}

Position-velocity diagrams for CS and SiO are presented in
Fig.~\ref{pv_diagram}. Note the symmetry around the X-axis in these
diagrams with low-velocity emission constituting a broad component of
weak emission. For both species, the high velocity component is more
pronounced toward the working surface of the outflow.

\begin{figure*}
\centering
\resizebox{\hsize}{!}{\includegraphics{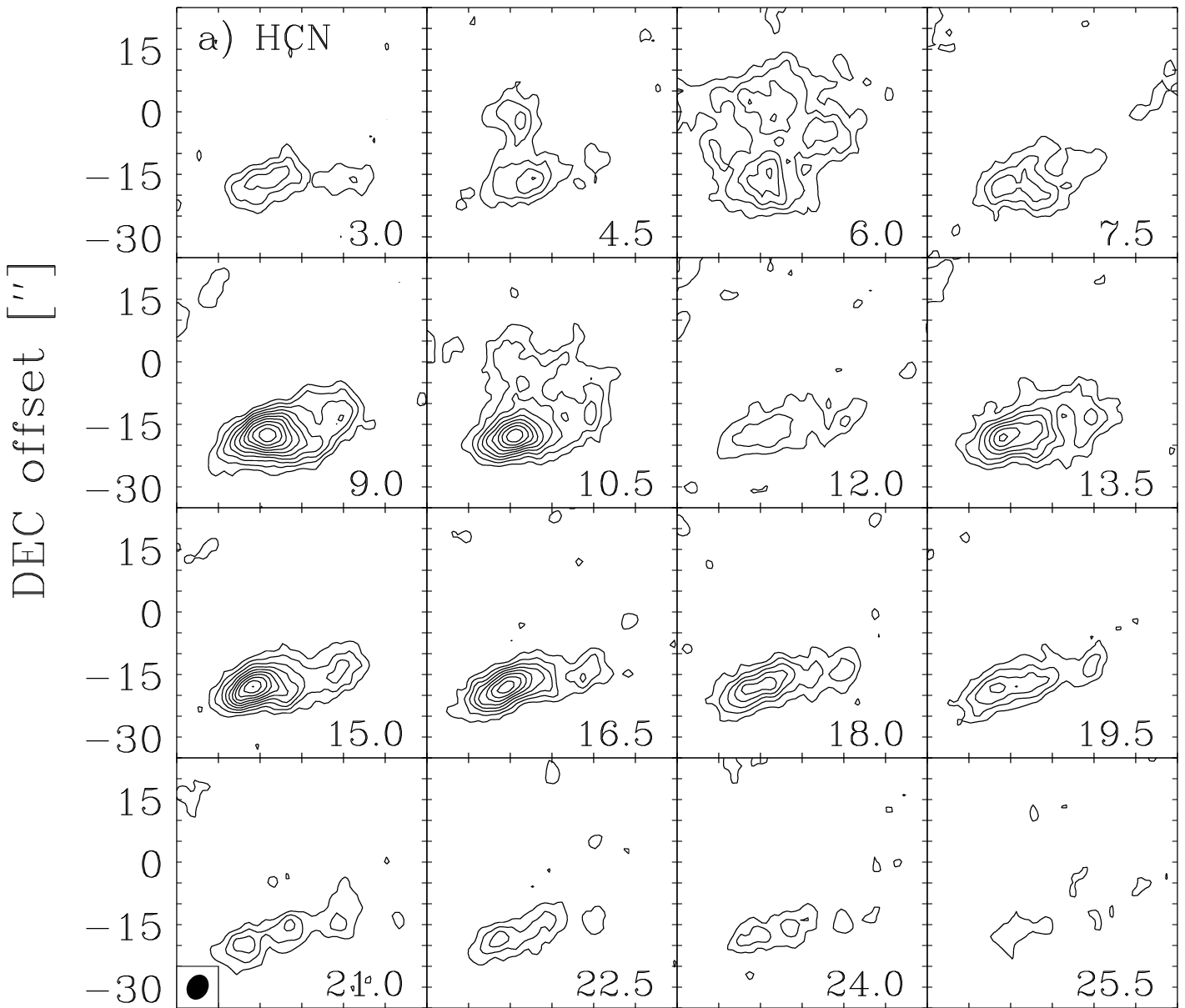}\includegraphics{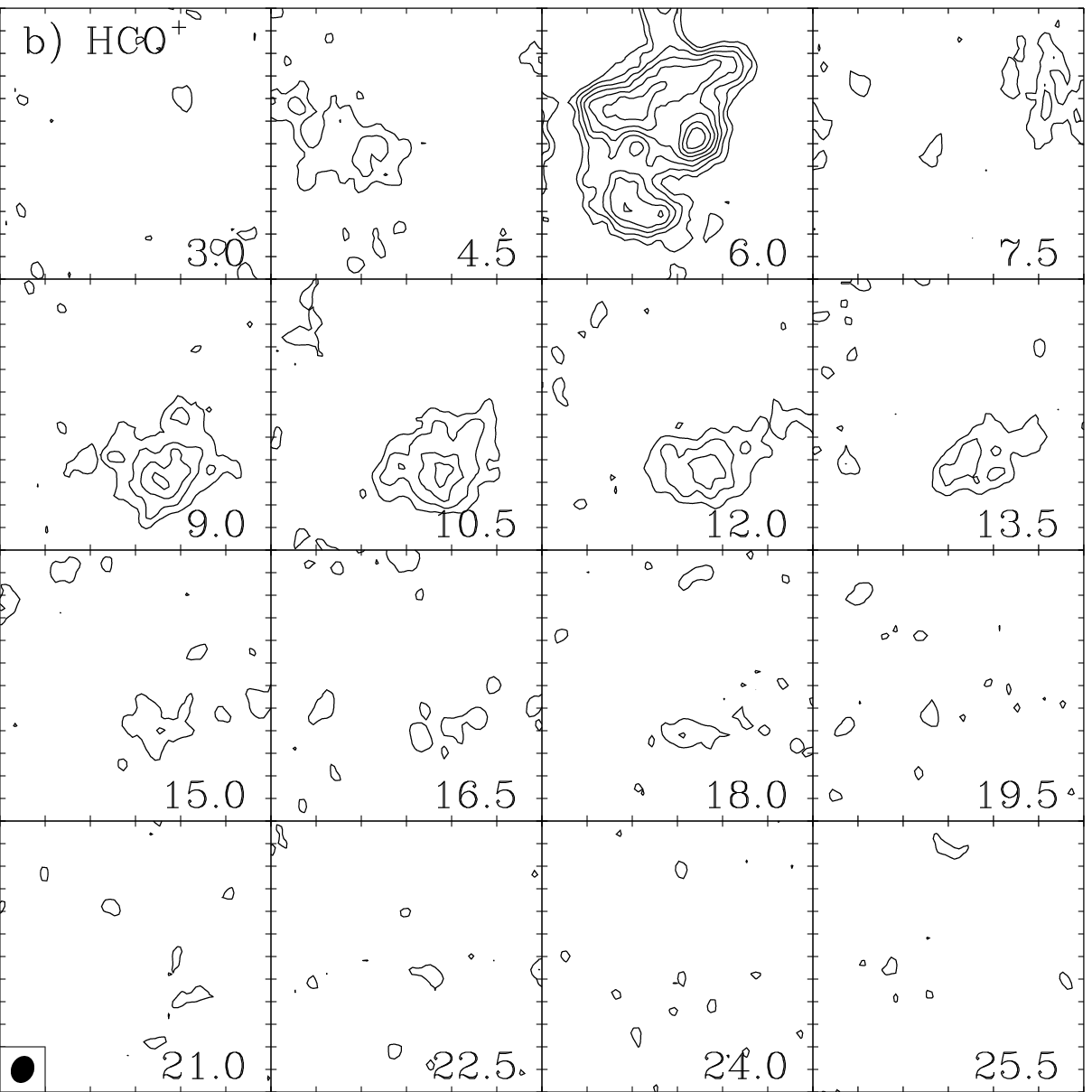}}
\resizebox{\hsize}{!}{\includegraphics{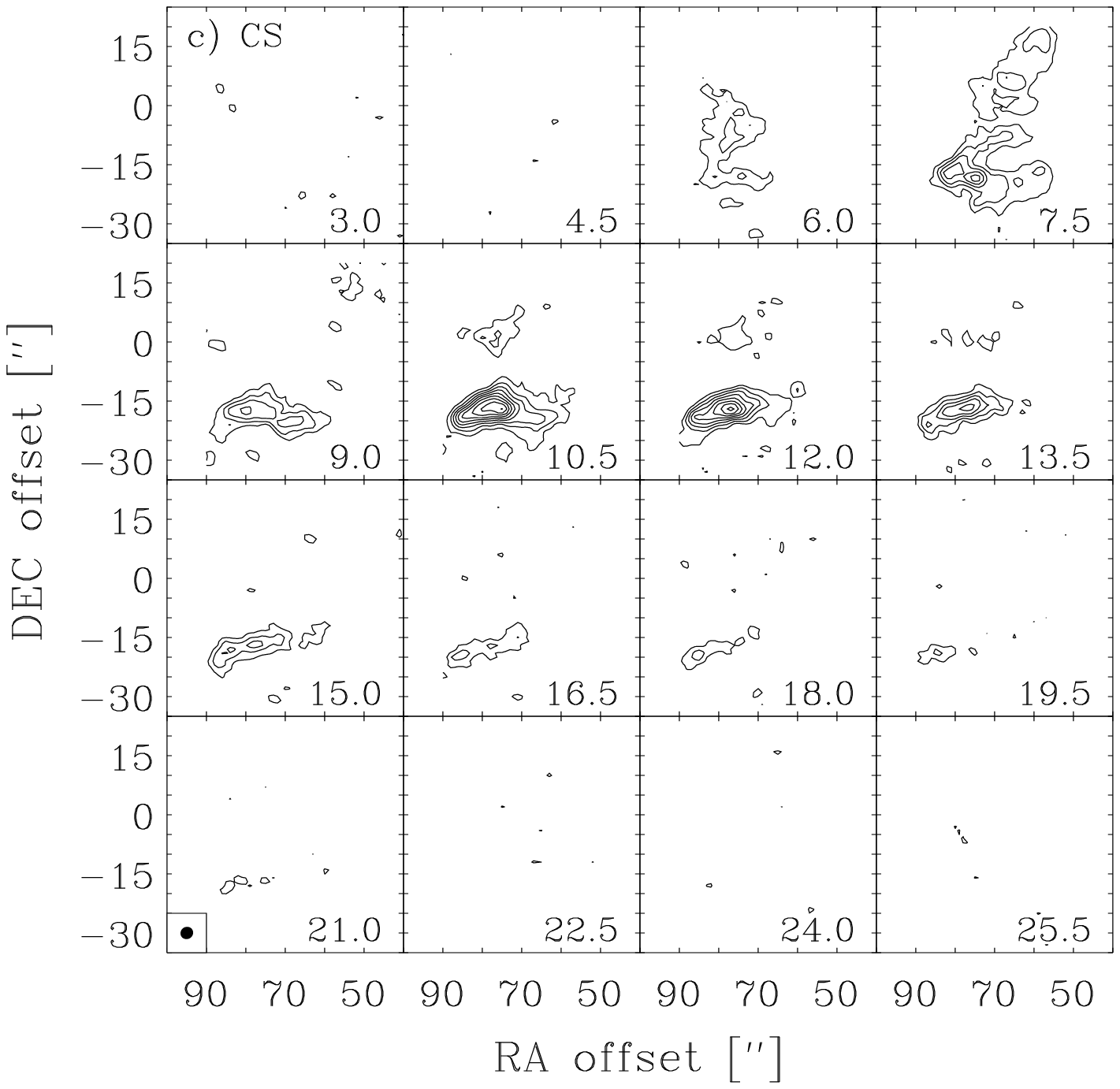}\includegraphics{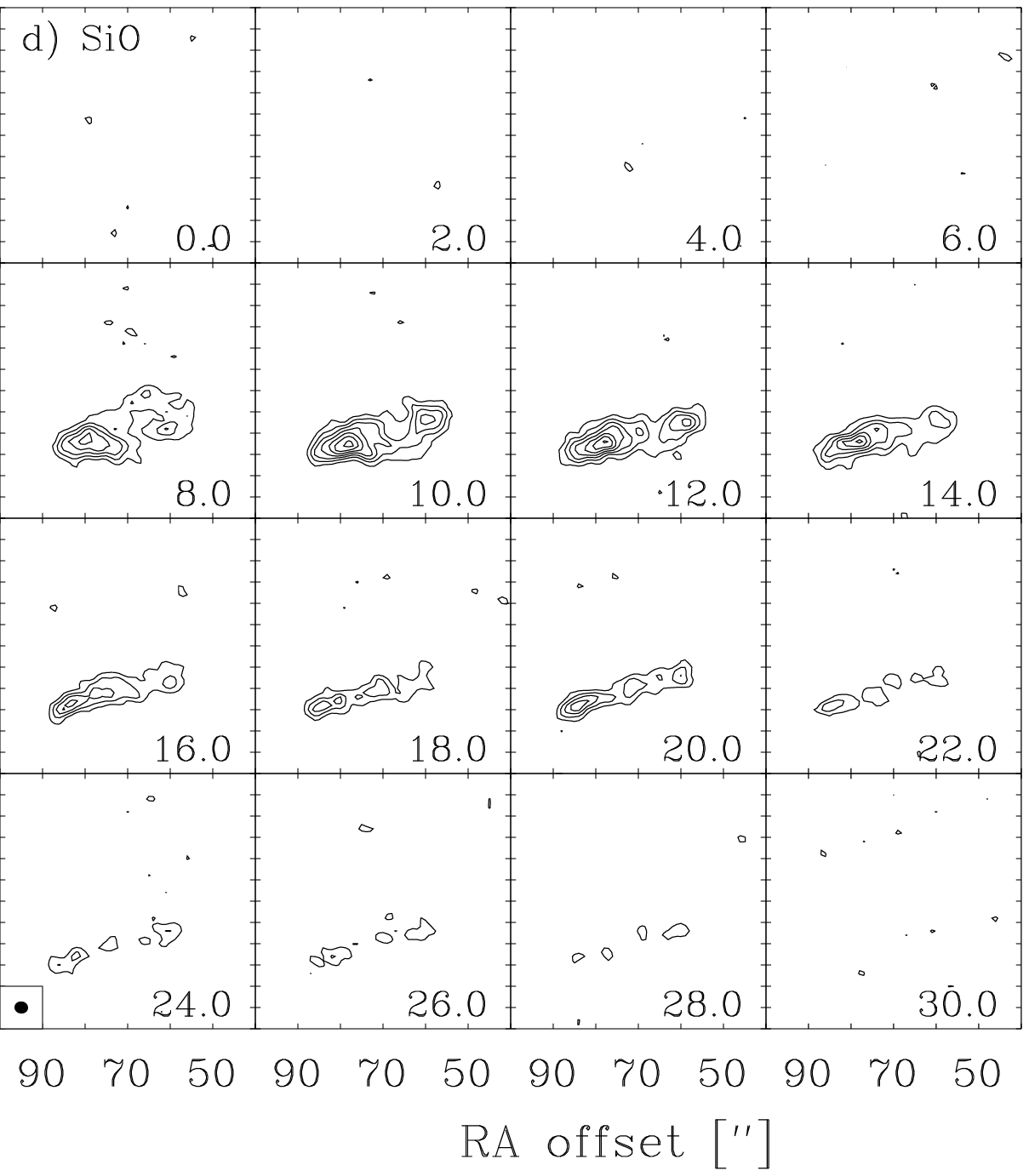}}
\caption{Channel maps for a) HCN, b) HCO$^+$, c) CS and d)
SiO. Contours are given in steps of 3$\sigma$. The synthesized beam is
indicated in the bottom left panel of each
figure.}\label{shock_cs_chan}
\end{figure*}

\subsection{Single-dish}\label{singledish}
\begin{figure*}
\resizebox{\hsize}{!}{\rotatebox{90}{\includegraphics{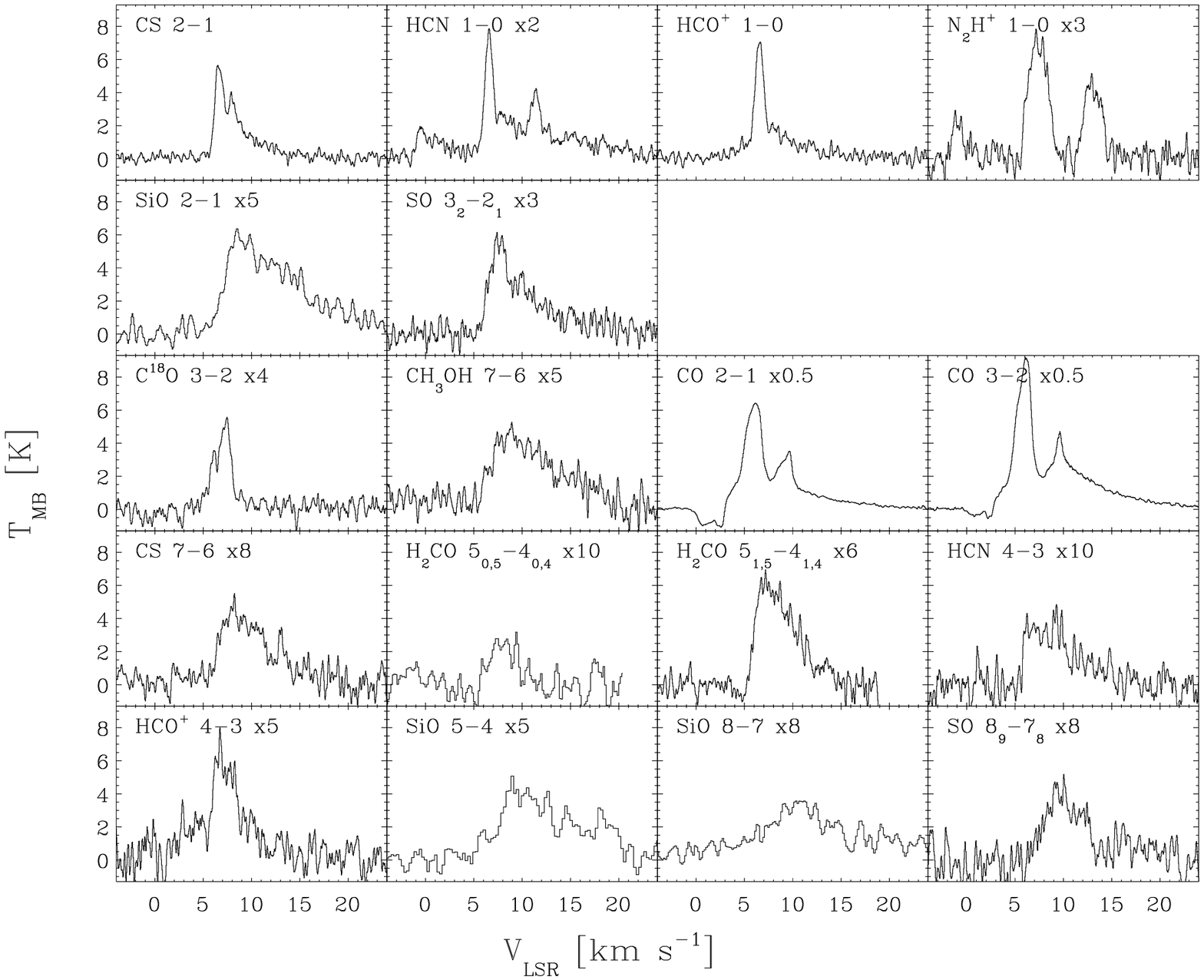}}}
\caption{Observed single dish spectra toward the IRAS2A shock
position. In the upper two rows the 3~mm observations from the Onsala
20~m are presented, whereas the 0.8--1.4~mm observations from the CSO
are shown in the lower 3 rows. All spectra are on the $T_{\rm MB}$
scale - note that some spectra have been scaled as indicated to fit on
the composite plot.}\label{overview_spectra}
\end{figure*}

The observed single-dish spectra are presented in
Fig.~\ref{overview_spectra}. In agreement with the interferometry
maps, the N$_2$H$^+$ hyperfine lines show Gaussian profiles with no sign
of outflow wings. The same applies for the C$^{18}$O 3--2
observations. The 3~mm lines of HCO$^+$, HCN, CS and SO show a ``two
component'' line profile with a narrow peaked profile close to the
systemic velocity of the cloud along with a clear red wing
extension. SiO on the other hand does not show the narrow component
but has a more or less abrupt increase slightly above the cloud rest
velocity with a close to linear decline in strength toward the higher
(red-shifted) velocities. For the CSO (0.8--1.4~mm) lines a similar
trend is seen (except for C$^{18}$O 3--2). For the CO 2--1 and 3--2
transitions two strong dips are seen at the cloud rest velocity and in
the blue part of the spectrum likely due to self-absorption and
possible switching onto non-source CO emission. This, however, does
not affect the wing emission in the spectrum, which stretches out to
$\sim 30$~km~s$^{-1}$ (20--25 km~s$^{-1}$ relative to the cloud rest velocity).

\begin{figure*}
\resizebox{\hsize}{!}{\rotatebox{90}{\includegraphics{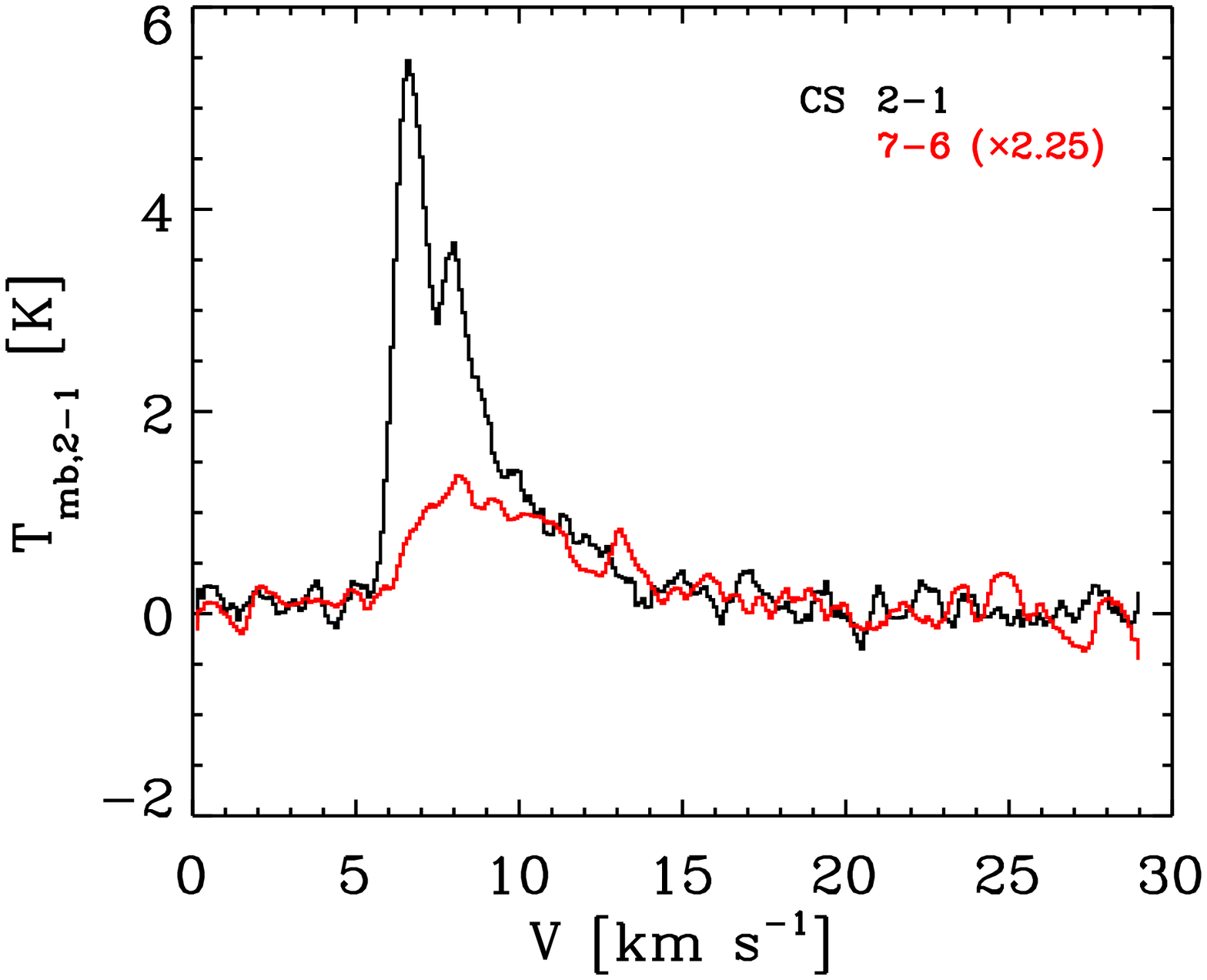}}\rotatebox{90}{\includegraphics{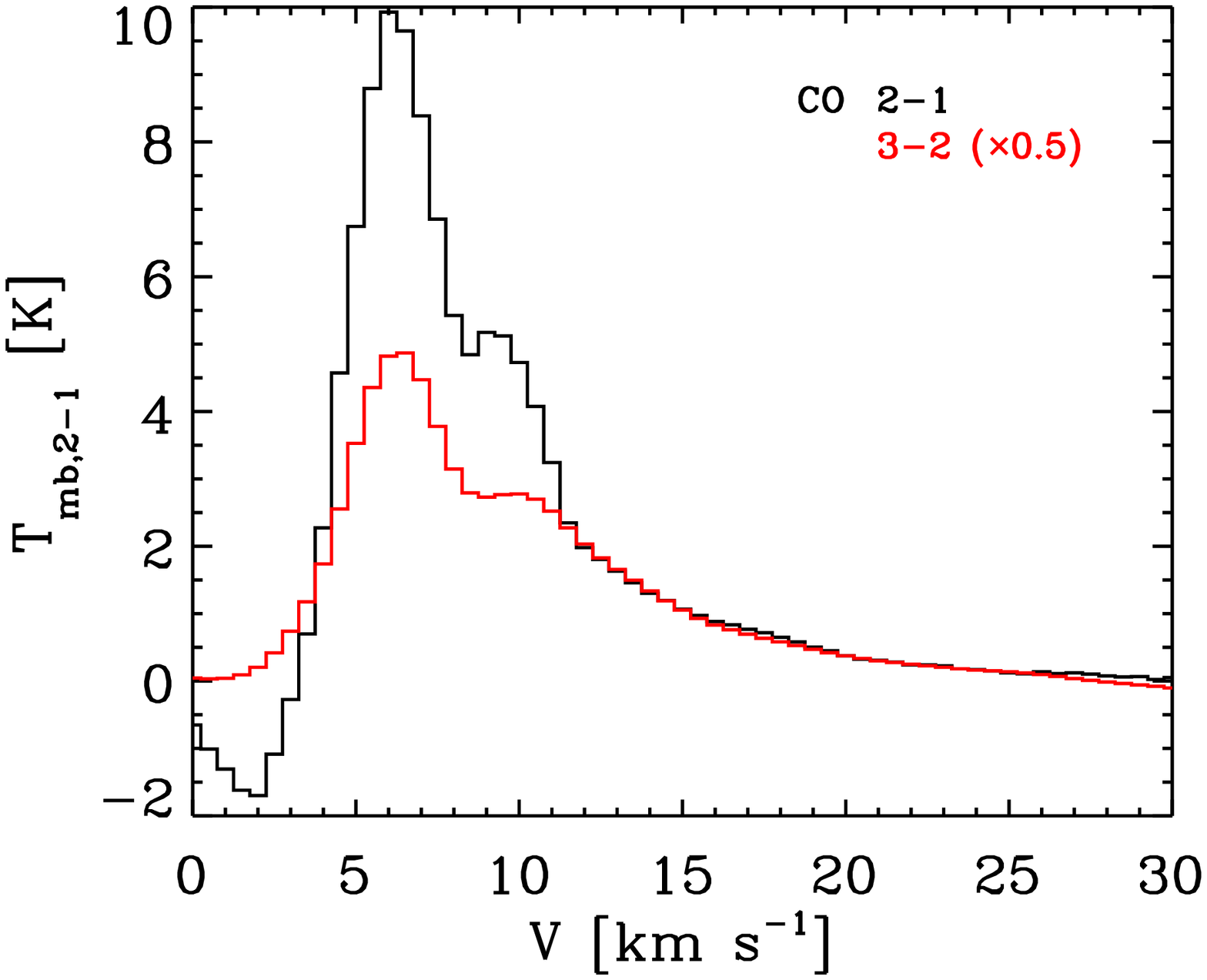}}}
\resizebox{\hsize}{!}{\rotatebox{90}{\includegraphics{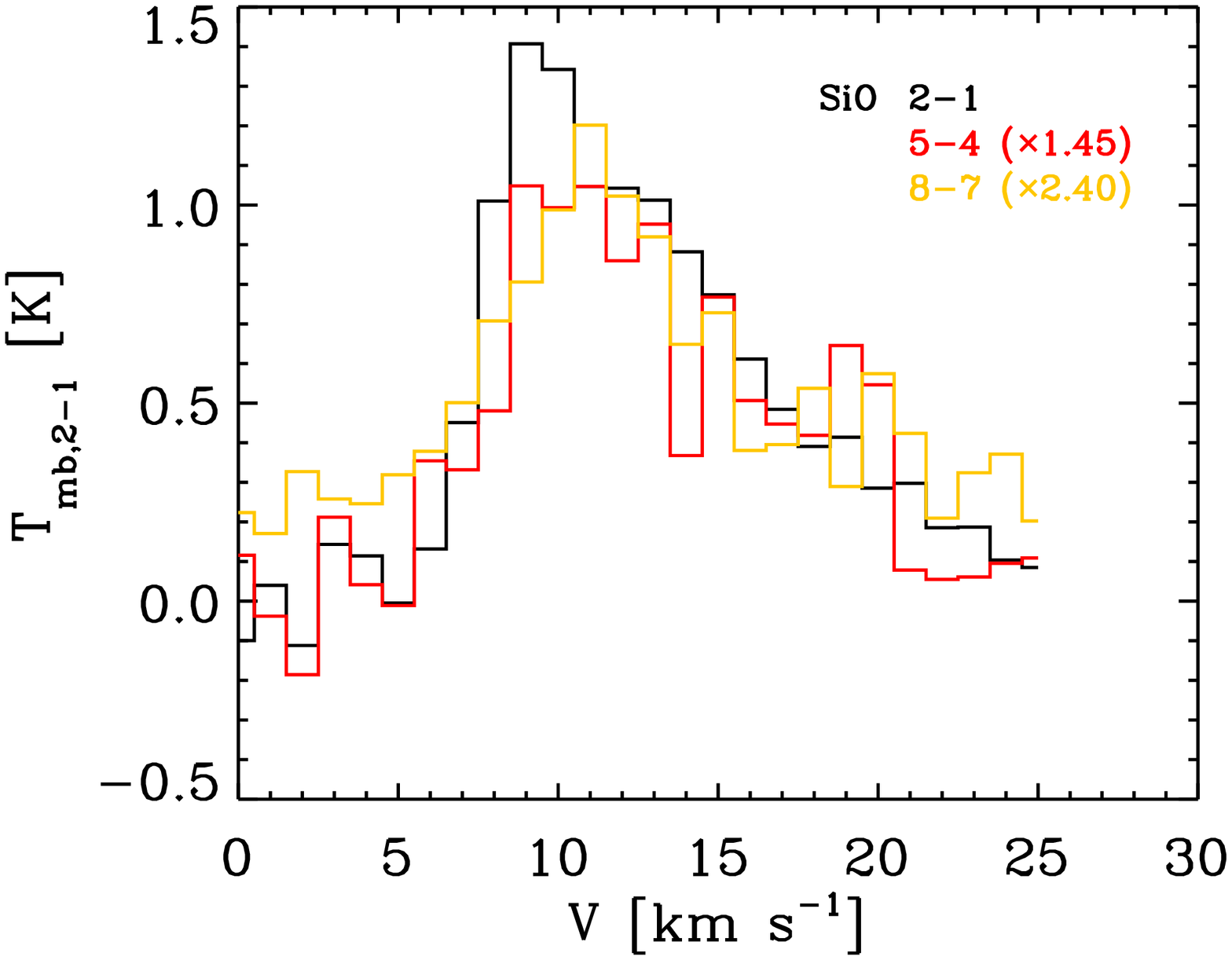}}\rotatebox{90}{\includegraphics{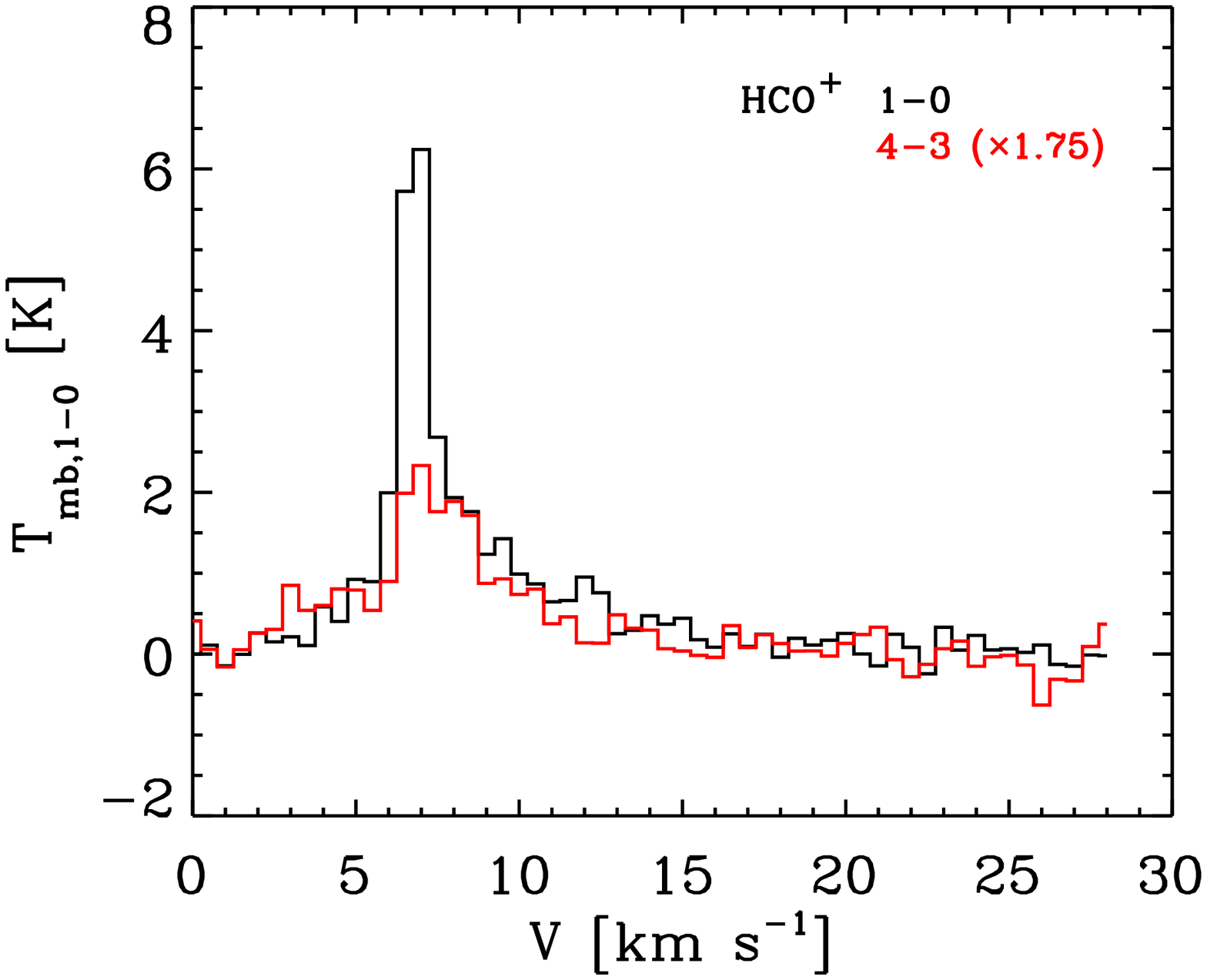}}}
\resizebox{\hsize}{!}{\rotatebox{90}{\includegraphics{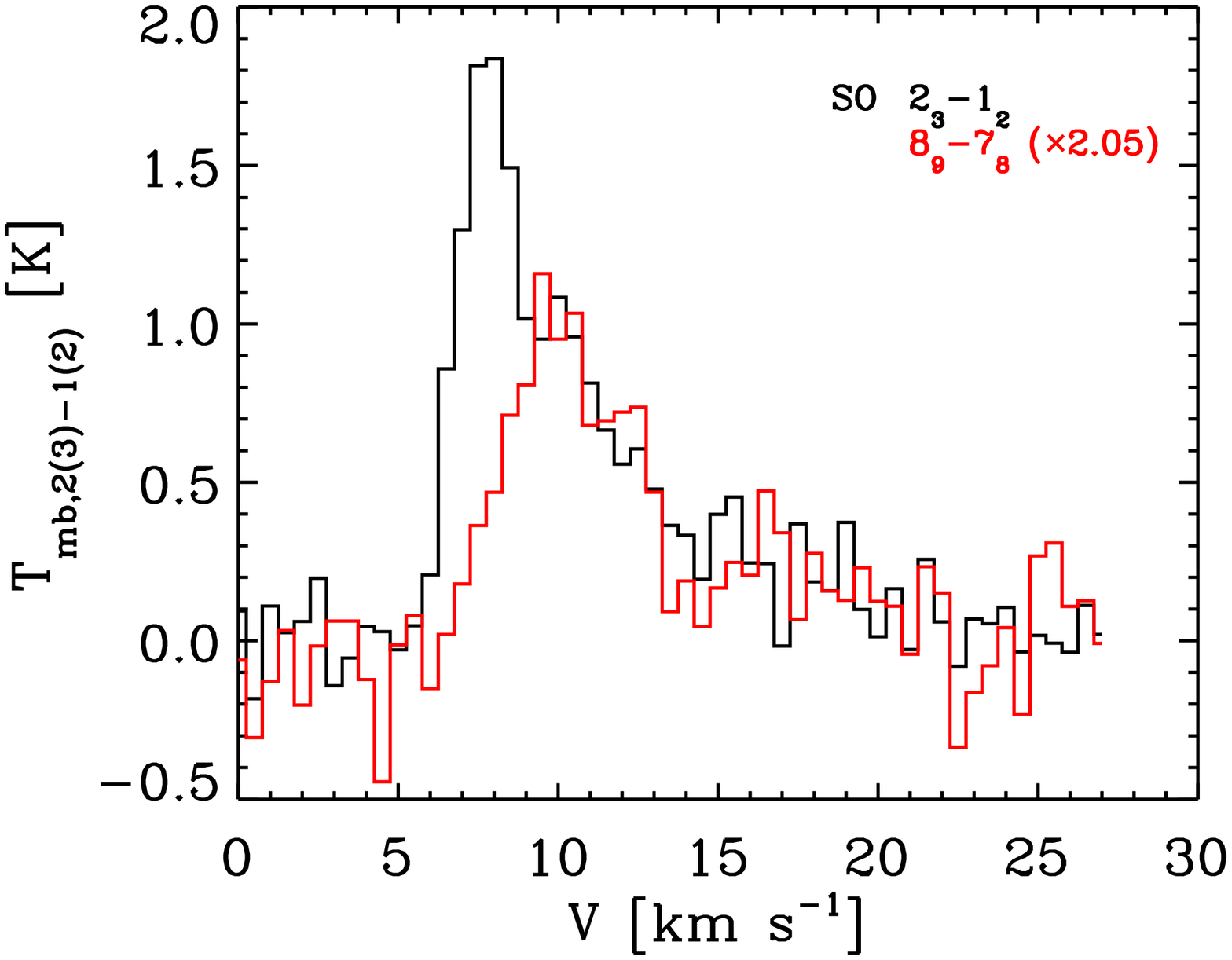}}\rotatebox{90}{\includegraphics{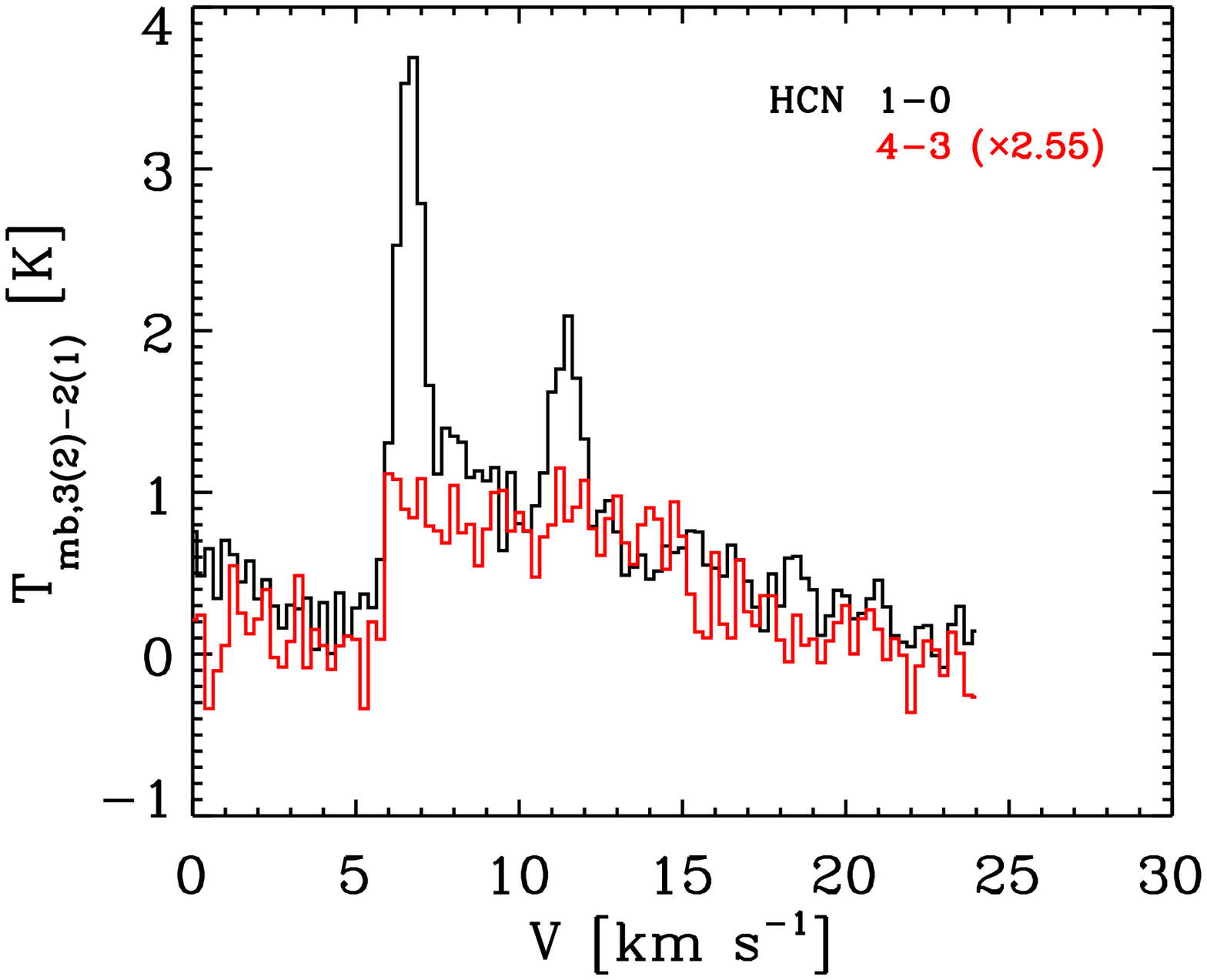}}}
\caption{Comparison between profiles for different transitions of
specific molecules. Note that the spectra have been smoothed by up to
a factor of 10 to bring out the agreement between the lines,
except for the scale factor given in the upper right corner of each
plot.}\label{lineratios}
\end{figure*}

In Fig.~\ref{lineratios} spectra for individual molecules are
compared. The two component separation into a core and wing profile
seems to be unique for each molecule: the widths of the core part of
the lines (where observable in more than one transition) and the
dependence of line strength with velocity in the wings, in particular
the terminal velocities inferred for each molecule, are both
independent of the observed transition. This gives a clear indication
that two distinct components with different excitation conditions and
chemical properties are observed and that the different molecules
probe distinct parts of each of these components, within which the
excitation properties do not vary significantly.

As for the interferometry data, the terminal velocities seem to
indicate different dynamical components of the outflow region. SiO and
CO probe material at the highest velocities relative to the cloud rest
velocity of 20--25~km~s$^{-1}$, whereas HCO$^+$ has the lowest degree of wing
emission extending $\lesssim$ 10~km~s$^{-1}$ from the rest velocity. The
remaining molecules fall somewhere in between. A similar trend was
seen in the \object{L1157} outflow by \cite{bachiller01}, who also
found CO and SiO to have significantly higher terminal velocities than
HCO$^+$, H$_2$CO, SO and CS. They suggested that this could be related
to differences in the formation mechanisms for the various
species. \citeauthor{bachiller01} also argued in favor of rather
homogeneous excitation conditions in the \object{L1157} outflow, since
maps of various transitions for specific molecules were found to be
very similar. In this case the main differences among different
molecular species would be the result of the chemistry in the region.

\subsection{Qualitative scenario}
The morphology of the interferometry maps and the line-profiles of the
single-dish observations can be explained in a simple picture as
signatures of a highly collimated outflow ramming into a quiescent
core or static cloud traced by the clumps of C$^{18}$O and N$_2$H$^+$
emission. The action of the outflow leads to sputtering of silicon off
dust grains to form SiO, which is totally absent in the quiescent
core. At the same time the abundances of CH$_{3}$OH, CS and SO increase,
more likely due to evaporation of ice mantle material. A sequence in
chemistry can be seen with SiO probing material at the highest
velocities followed by CS and SO, thought to be a result of enhanced
sulfur gas phase chemistry, and CH$_{3}$OH, H$_2$CO and HCN, which are
likely to be direct results of grain mantle release. The difference in
extent of the line wings indicates either that the molecule formation
time scales are varying, with SiO being produced most rapidly in the
shock, or that the more volatile species such as CH$_{3}$OH do not
survive at the highest velocities in the flow.

That N$_2$H$^+$ is observed only in the quiescent cloud material is
explained if the temperature in the material affected by the outflow
increases to $\gtrsim 20$~K. At this temperature CO is released from
grain mantles and becomes the dominant destruction channel of N$_2$H$^+$,
lowering the abundance of this molecule. Comparison between the
morphology of the N$_2$H$^+$ emission and that of other species indicates
a clear interaction between the outflowing material and the ambient
cloud. Fig.~\ref{ovro_moments}-\ref{bima_moments} show a cavity of
N$_2$H$^+$ emission where the shocked gas (e.g., SiO) appears. The two
peaks seen in SiO, HCN and CS also seem to be related to an increase
in N$_2$H$^+$ emission. A natural question is whether the outflow shapes
or is being shaped by the ambient material. Judging from the
morphology of the larger scale emission in Fig.~\ref{scuba_overview},
it is striking to note the presence of large amounts of dust northeast
of the central \object{IRAS2A} protostar. The two CO outflows
identified by different authors
\citep[e.g.,][]{liseau88,engargiola99,knee00} trace the edges of this
dust condensation. The two perpendicular outflows seen toward IRAS2
may therefore reflect the conditions in the ambient cloud material
rather than the intrinsic properties of the central protostellar
system: the CO outflow could simply be deflected around the dense
material traced by the N$_2$H$^+$ and continuum emission leading to the
quadrupolar morphology. In either case, however, this does not change
the interpretation of the shocked material in this paper. High
velocity gas ($\gtrsim$~20~km~s$^{-1}$ relative to the systemic velocity) is
present toward the eastern lobe as it is seen from most of the species
observed in this paper and this indicates the presence of the shock.

\section{Analysis}\label{analysis}
\subsection{Line intensities}
Despite the clear separation between the ``quiescent'' and ``shocked''
parts of the line profiles for most molecules, the disentanglement of
the emission into a core and wing component is not unique. Each line
was decomposed in two parts, with the emission integrated over
velocities higher and lower (wing and core components, respectively)
than where the profiles in Fig.~\ref{lineratios} separate. For C$^{18}$O
and N$_2$H$^+$ Gaussians were fitted to each line giving widths of
1.5--2~km~s$^{-1}$ (FWHM), which compares well to the widths of the core
parts of the remaining lines. Table~\ref{line_intens} lists the
resulting line intensities.

\begin{figure}
\resizebox{\hsize}{!}{\includegraphics{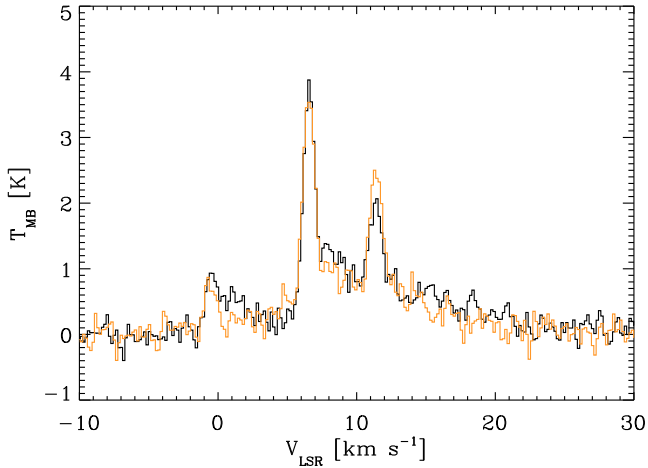}}
\caption{The HCN $1-0$ spectrum toward the shock position (black)
overplotted with a composite of three versions of the HCO$^+$ spectrum
toward the same position (red) - each shifted with the measured
shifts between the HCN hyperfine lines and scaled according
(relatively) to the 1:3:5 line-ratio expected for optically thin
emission from material in LTE.}\label{hco_hcn}
\end{figure}

The hyperfine splitting of the HCN 1--0 line gives rise to three
components within the same setting. The two weaker transitions are
offset $-5$ and 7~km~s$^{-1}$ relative to the main hyperfine line. Overlap
in the line wings therefore makes the interpretation of this line
difficult. If the emitting material is in local thermodynamical
equilibrium (LTE) and the emission is optically thin, one should
expect the hyperfine components to be in a ratio of 1:3:5. In
Fig.~\ref{hco_hcn} a comparison between the HCO$^+$ and HCN spectra
toward the shock position is shown. The HCN spectrum has been
overplotted with a composite of three versions of the HCO$^+$ spectrum
shifted according to the frequency shifts of the hyperfine lines and
scaled in the relative 1:3:5 proportions - leaving an overall
``normalization factor'' between the HCN and HCO$^+$ line intensities
as the only free parameter. The good agreement is remarkable and
indicates that HCO$^+$ and HCN trace the same material, especially at
the lower velocities. The intensity of the main hyperfine component of
the HCN 1--0 line is thereby found to be 0.5 times the intensity of
the HCO$^+$ 1--0 line. It should still be re-emphasized, that the
interferometer maps show clear differences for the less extended HCN
and HCO$^+$ emission at higher velocities.

\begin{table*}
\caption{Line parameters from single dish observations}\label{line_intens}
\begin{tabular}{llllll}\hline\hline
Molecule & Transition & $V({\rm core)}$$^{a}$ & $I({\rm core})$$^{b}$ & $V({\rm wing)}$$^{a}$ & $I({\rm wing})$$^{b}$ \\
          &               & km~s$^{-1}$       & K km~s$^{-1}$         & km~s$^{-1}$           & K km~s$^{-1}$         \\ \hline
$^{12}$CO & 2--1          & [4,11]      & 45              & [11,30]         & 13              \\
          & 3--2          & [4,11]      & 50              & [11,30]         & 24              \\
C$^{18}$O & 3--2          & Gaussian    & 1.7             & $\ldots$        & $\ldots$        \\
CH$_3$OH  & 7$_2$--6$_2$  & [6,9]       & 2.0             & [9,21]          & 6.4             \\
CS        & 2--1          & [5.5,9.5]   & 12              & [9.5,16]        & 3.9             \\
          & 7--6          & [5.5,9.5]   & 1.6             & [9.5,16]        & 1.6             \\
HCN       & 1--0$^{c}$    & [5.5,7.5]   & 4.2             & [7.5,15]        & 3.3             \\
          & 4--3          & [5.5,7.5]   & 0.72            & [7.5,15]        & 0.78            \\
HCO$^+$   & 1--0          & [5.5,7.5]   & 8.4             & [7.5,15]        & 6.5             \\
          & 4--3          & [5.5,7.5]   & 2.7             & [7.5,15]        & 2.0             \\
H$_2$CO   & $5_{1,5}-4_{1,4}$ & [5.5,7.5]   & 3.1         & [7.5,15]        & 1.5             \\
SiO       & 2--1          & $\ldots$    & $\ldots$        & [7,29]          & 10.4            \\
          & 5--4          & $\ldots$    & $\ldots$        & [7.5,20]        & 6.5             \\
          & 8--7          & $\ldots$    & $\ldots$        & [7.5,28]        & 4.4             \\
SO        & $2_3-1_2$     & [5,9]       & 4.3             & [9,17]          & 4.2             \\
          & $8_9-7_8$     & [5,9]       & 0.61            & [9,17]          & 2.0             \\ \hline
\end{tabular}

Notes: $^{a}$Velocity interval over which the emission is
integrated. $^{b}$Integrated line intensity ($I=\int T_{\rm mb}\, {\rm
d}v$). $^{c}$Main hyperfine line; line intensities derived through
decomposition with HCO$^+$ line. See description in text.
\end{table*}

\subsection{Tying interferometry and single-dish observations together}
One question to address is whether the single-dish and interferometry
observations trace the same material. The similar trends seen in the
data from the two types of observations seem to support this, but it
is also known that the interferometry observations lack sensitivity to
extended structures on larger scales.

\begin{figure}
\resizebox{\hsize}{!}{\includegraphics{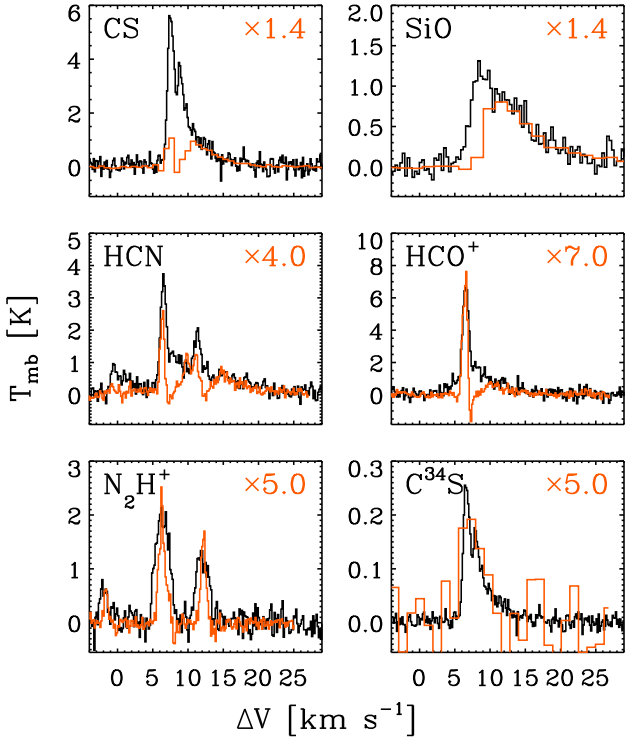}}
\caption{Comparison between the single-dish observations (black) and
corresponding spectra from the interferometer observations restored
with the single-dish beam (red) at the shock position. The spectra
from the interferometry observations have been scaled to resemble the
wing of the single-dish spectra with the factors indicated in the
upper right corner. For C$^{34}$S the single-dish CS has been
downscaled by a factor 22.}\label{resolved_out}
\end{figure}

Fig.~\ref{resolved_out} presents comparisons between the single-dish
observations and spectra taken from interferometry data sets, restored
with beam sizes appropriate for the Onsala 20~m telescope. The
interferometry spectra have been scaled to match the wings of the
single-dish spectra. Only a small correction of the order of 30\%
needs to be applied to match the CS and SiO spectra, which is close to
the calibration uncertainty. This confirms that the higher velocity
emission is relatively compact (as the maps also suggest for the
emission in the transverse direction of the outflow), making it less
subject to the incomplete $(u,v)$ sampling of the interferometer
observations.

Closer to the systemic velocity of the cloud ($\approx 7$~km~s$^{-1}$) the
discrepancy between the single-dish and interferometry spectra
increases. The CS interferometry observations pick up only a small
fraction of the emission in the ``core'' part of the single-dish
line. The dip seen in the single-dish CS spectra at the rest velocity
of the cloud is a result of self-absorption, while for the
interferometry observations it is caused by the interferometer
resolving out extended emission close to the cloud systemic
velocity. The $(u,v)$ sampling is also responsible for the lack of
emission in the SiO interferometry spectra at velocities close to the
rest velocity, although it is less significant for this molecule.

For the BIMA observations of HCN and HCO$^+$, emission close to the
systemic velocity is still mostly resolved out as indicated by the
dips in the interferometer HCN and HCO$^+$ spectra and as seen in the
channel maps in Fig.~\ref{shock_cs_chan}. The slightly better $(u,v)$
coverage from BIMA, however, makes these lines less subject to
resolving out at velocities different from the systemic velocity.

Nyquist sampled single-dish maps of the different molecular
species would make it possible to combine the interferometer and
single-dish data to create maps including the short-spacings. This
would settle the issue of the differing $(u,v)$ coverage of the two
arrays. In this paper we only have single pointing observations. The
agreement between the single-dish and interferometry spectra in the
line wings, however, justifies the discussion of the outflow
component based on the morphology in the interferometry maps presented
in Sect.~\ref{interferometry}. The agreement also makes it possible to
use the interferometer maps to determine the spatial extent of the
wing component, and thereby to estimate the beam filling factor for
the single-dish observations. The wing part of the SiO and CS
interferometry maps give a rough estimate of the extent of the outflow
emission in the transverse direction of 5-10\arcsec, leading to
filling factors ranging from 0.07 to 0.32 for the single-dish beam
sizes. For the core component of the lines on the other hand, the
interferometer observations are less useful because of the significant
fraction of the low-velocity, extended emission that is resolved
out. For the following discussion, a filling factor of unity is
therefore assumed for the core component of the single-dish data,
which seems realistic as the interferometry maps do reveal emission
extended over scales larger than 20--30\arcsec.

\subsection{Statistical equilibrium calculations}\label{stateq}
In order to derive the physical properties of the emitting gas and the
column densities of the various molecules, one-dimensional statistical
equilibrium calculations were performed using the 1D radiative
transfer code called Radex. Radex uses the escape
probability formalism to solve the statistical equilibrium equations
for a medium with constant density and temperature
\citep[see][]{jansen94,schoeier03radex}. For subthermally excited
emission this approach is an improvement compared with the rotation
diagram method, as demonstrated for interpretations of the CH$_{3}$OH
emission toward the \object{IRAS2A} outflow by \cite{bachiller98}. The
statistical equilibrium calculations also treats opacity effects in
the correct way, again contrasting the rotation diagram analysis which
relies on optically thin emission.

For each molecule, line intensities were calculated for varying column
density, density of the main collision partner (H$_2$) and kinetic
temperature. The filling factors estimated on the basis of the
interferometry observations as discussed above were adopted and the
calculated line intensities were compared to the observed ones. Since
our main interest is in the relative behavior of the lines some of the
uncertainties in the assumptions, e.g., the filling factor of unity
for the core component, will cancel out, if the differences between
the observed lines are not too large.

\begin{figure}
\resizebox{\hsize}{!}{\includegraphics{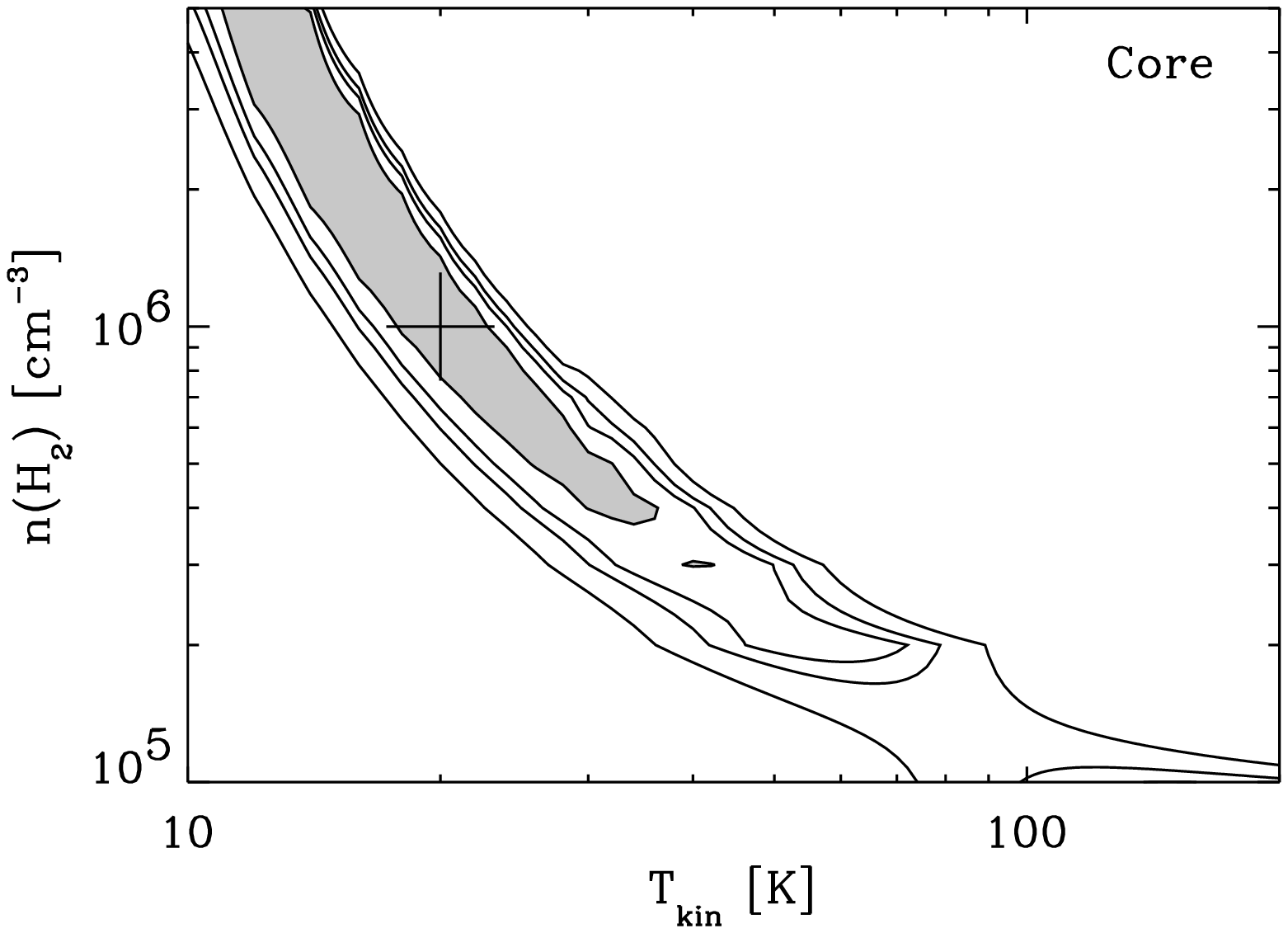}}
\resizebox{\hsize}{!}{\includegraphics{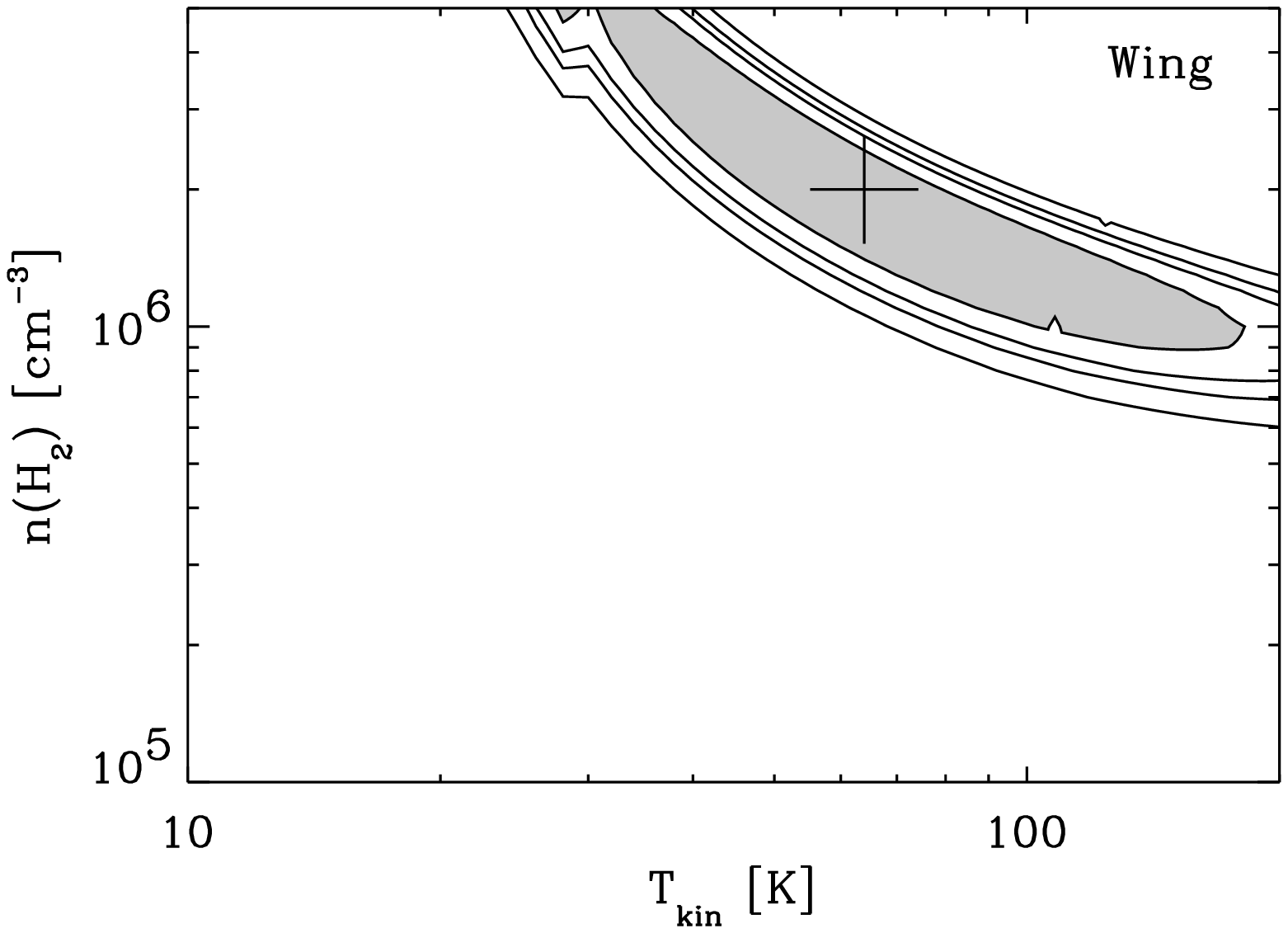}}
\caption{Best fitted densities and kinetic temperatures derived from
statistical equilibrium calculations with the assumptions described in
the text. In the upper panel the results for the core part of the
lines are shown and in the lower panel results for the wing part. The
grey-scaled contour indicate the 1$\sigma$ confidence region, while
the other contours, indicate 2$\sigma$, 3$\sigma$ and 4$\sigma$
confidence levels. For each plot the column densities given in
Table~\ref{radex_results} have been used for creating the cut in the
cube of models with varying temperature, density and column
density. For each plot the black `+' indicate the best-fit values. The
reduced $\chi^2$ is 1.8 (8 fitted lines) for the core component and
3.8 (13 fitted lines) for the wing component.}\label{radex_lineratios}
\end{figure}

The comparison with the observations was performed by calculating the
$\chi^2$-statistics for each set of parameters. The uncertainty in the
derived line intensities due to the calibration and the
disentanglement of the core and wing components was assumed to be
30\%. The best fit models for the different species agree quite well
in the $(T_{\rm kin},n_{\rm H_2})$ plane, so all lines are combined
into a single $\chi^2$ estimate to constrain the parameters. This is
illustrated in Fig.~\ref{radex_lineratios} where the constraints on
densities and temperatures have been plotted for given (optimal)
values of the column densities for the individual molecules. For the
core component an H$_2$ density of $\sim 10^6$~cm$^{-3}$ and
temperature of 20~K is found to be consistent with the observations
with a reduced $\chi_{\rm red}^2$ of 1.8 for 8 fitted lines. Note that
the density and temperature are closely coupled, making individual
determinations somewhat ambiguous, as illustrated in the $\chi^2$
plot, where it is seen that a lower temperature and correspondingly
higher density are equally probable. A lower density/higher
temperature can also not be completely ruled out. For the wing
component a best fit density of 2.0$\times 10^6$~cm$^{-3}$ and
temperature of 70~K is found with $\chi_{\rm red}^2$ of 3.8 for 13
fitted lines. The temperature is slightly lower than the value quoted
by \cite{bachiller98}, but still within the mutual uncertainties. The
derived column densities assuming these temperatures and densities are
given in Table~\ref{radex_results}.

\begin{figure}
\resizebox{\hsize}{!}{\rotatebox{90}{\includegraphics{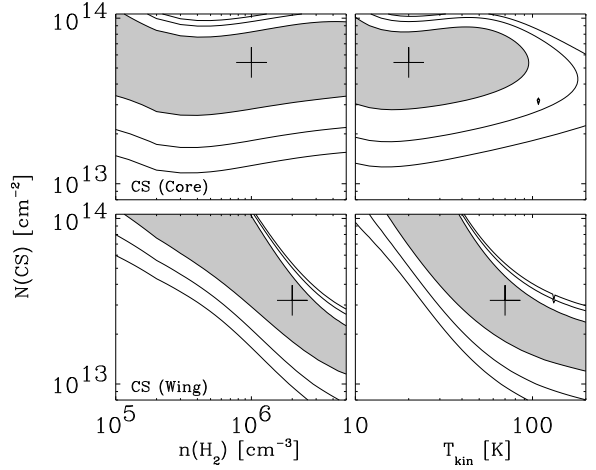}}}
\caption{Confidence plots for column density vs. kinetic temperature
(right) and density (left) for the CS observations of the core
component of the line profile (upper panels) and wing part (lower
panels). The contours correspond to the 1$\sigma$ (solid grey),
2$\sigma$ and 3$\sigma$ confidence levels. The derived column
densities are well constrained for the set of temperatures and
densities determined from Fig.~\ref{radex_lineratios} (black
`+').}\label{coldens_conf}
\end{figure}

The kinetic temperatures and densities may vary between regions traced
by different molecular species. Both the velocity profiles of the
various molecules and the structure of their emission in the
interferometry maps show dissimilarities, indicating chemical
differentiation possibly due to a combination of the shock evolution
and variations in the physical conditions. On the other hand as
illustrated in Fig.~\ref{lineratios}, the conditions for lines of a
particular molecule are remarkably homogeneous over the entire shock
velocity range. It is also found that the derived parameters do not
vary much when specific molecules are included or not. The constraints
put on our derived temperature and density are in rough agreement with
measurements from other shocked regions from molecular outflows
\citep[e.g.,][]{bachiller97,garay98}. As a first order approximation
the two component structure therefore seems to describe the excitation
conditions well. Moreover, the derived column densities are well
constrained with the assumed temperature and density in this regime,
as illustrated in Fig.~\ref{coldens_conf}.

\begin{table}
\caption{Column densities for the various molecules from statistical
equilibrium calculations.}\label{radex_results}
\begin{tabular}{lll}\hline\hline
Molecule & $N({\rm core)}$$^{a}$ & $N({\rm wing})$$^{a}$ \\
                & [cm$^{-2}$]  & [cm$^{-2}$] \\ \hline
CO$^{b}$        & 5.1$\times 10^{17}$  & 1.2$\times 10^{17}$ \\ 
CH$_3$OH        & 3.4$\times 10^{15}$  & 5.8$\times 10^{15}$ \\ 
CS              & 5.4$\times 10^{13}$  & 3.2$\times 10^{13}$ \\
HCN             & 1.1$\times 10^{13}$  & 5.7$\times 10^{12}$ \\ 
HCO$^+$         & 6.6$\times 10^{12}$  & 3.5$\times 10^{12}$ \\ 
H$_2$CO         & 8.9$\times 10^{13}$  & 1.8$\times 10^{13}$ \\ 
N$_2$H$^+$      & 9.0$\times 10^{12}$  & $\ldots$    \\
SiO             & $\ldots$     & 1.4$\times 10^{14}$ \\
SO              & 6.2$\times 10^{13}$  & 1.3$\times 10^{14}$ \\ \hline
\end{tabular}

Notes: $^{a}$Calculated for $n_{\rm H_2} = 1\times 10^6 {\rm cm}^{-3}$
and $T=20$~K (core) and $T=80$~K (wing). $^{b}$Based on $^{12}$CO
measurements for the column densities in the wings and C$^{18}$O
measurements for the core (assuming a $^{16}$O/$^{18}$O isotope ratio
of 540).
\end{table}

\section{Discussion}\label{discussion}
\subsection{Comparison to other protostellar outflows and envelopes}
As mentioned in Sect.~\ref{introduction} only a few studies have
addressed the chemistry in outflow regions in detail, the most
thorough being that of the \object{L1157} outflow
\citep[e.g][]{bachiller97,bachiller01} and the molecular condensation
ahead of the \object{HH2} outflows \citep{girart02}. These regions
differ significantly in their context: the CO maps \object{L1157} show
an outflow progressing through the protostellar envelope and extended
cloud. This causes a number of bow shocks and condensations of
enhanced density and varying chemistry along the outflow axis. The
chemistry in \object{HH2} is thought to be induced by irradiation of
the molecular condensation through UV flux from the bright Herbig-Haro
object. In such regions especially HCO$^+$ and NH$_3$ should be
greatly enhanced according to the models of \cite{viti99outflow}.

Table~\ref{abundtable} lists the abundances found for the two
components of the \object{IRAS2A} outflow, calculated as simple ratios
between the column densities - taking the CO abundance relative to
H$_2$ at constant value of $1\times 10^{-4}$. This means that the
quoted abundances for the ``wing component'' are averaged over
material with a large range of velocities. As indicated by the maps
and the single-dish line profiles the emission from some of the
molecules may be more concentrated toward material with lower
velocities. In this sense, the quoted abundances are therefore lower
limits to the abundances in the regions where these molecules are
observed.

Table~\ref{abundtable} also compares the derived abundances to
those found for the \object{L1157} and \object{BHR71} outflows
\citep{bachiller97,garay98}, the molecular condensation ahead of the
HH2 object \citep{girart02} and other types of protostellar
environments - in particular the IRAS2A protostellar envelope
\citep{paperii}, the ``hot component'' of the IRAS~16293-2422 envelope
\citep{schoeier02} and the the ``C'' position of the dark cloud L134N
\citep{dickens00}. The abundances in the IRAS~16293-2422 envelope were
derived through detailed radiative transfer of the dust continuum and
molecular line data. The CO, CS and HCN abundances quoted are averages
over the entire envelope. Since the abundances in the outer part of
the envelope may be lower due to freeze-out (see for example
discussion for CO in \cite{jorgensen02}), the quoted numbers are
likely to be lower limits to the abundances in the inner, warm region
of the envelope.

\begin{figure}
\resizebox{\hsize}{!}{\includegraphics{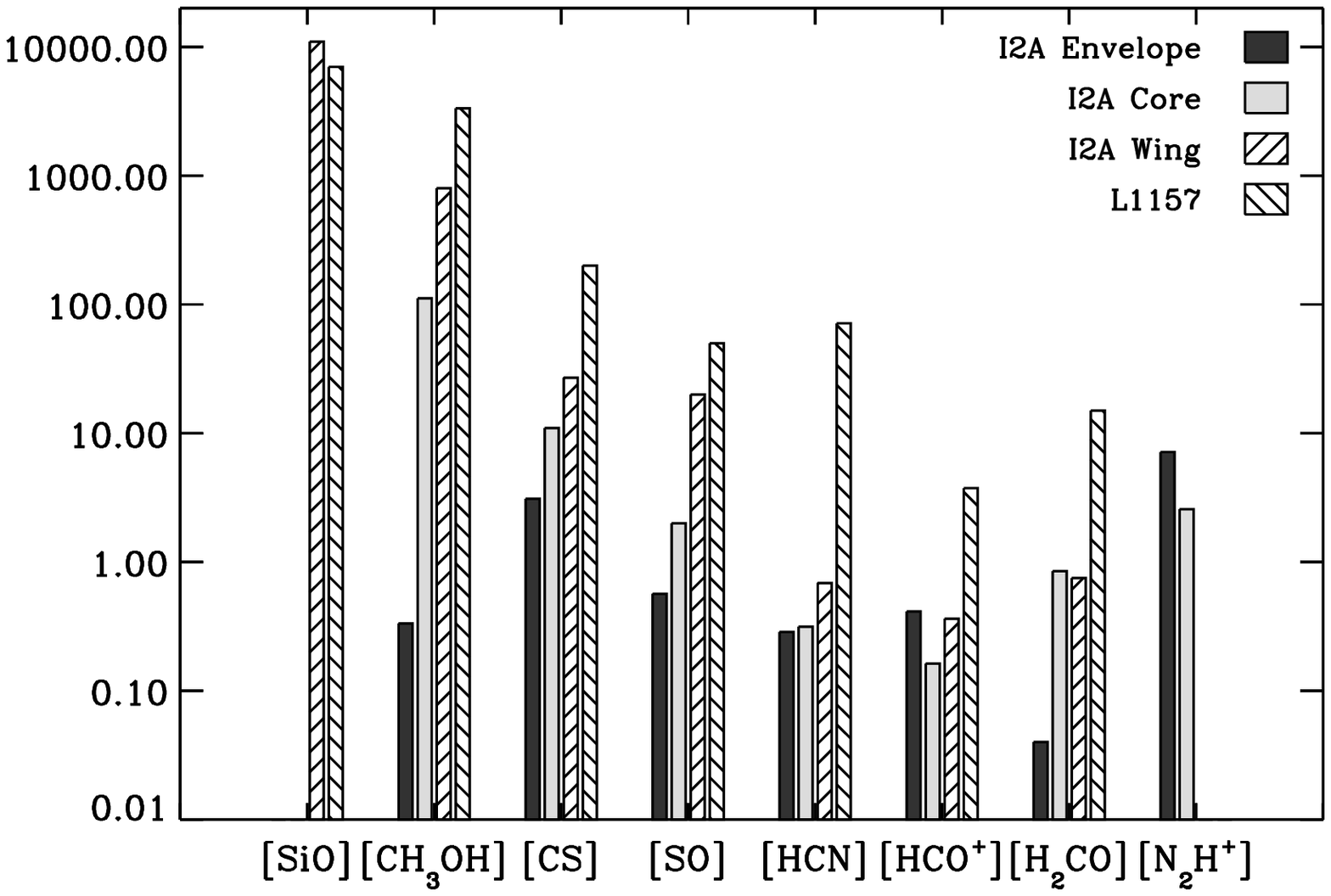}}
\caption{Abundances in the IRAS2A envelope \citep{paperii} compared to
the core and wing position components of the IRAS2A outflow (this
paper) and the L1157 outflow \citep{bachiller97}. The abundances have
been normalized to the abundances of L134N. For SiO, which has not
been detected in L134N, a ``reference'' abundance of $10^{-11}$ has
been assumed. Note that N$_2$H$^+$ is not detected in the outflows,
whereas SiO is not detected in the quiescent
components.}\label{abundfig}
\end{figure}

In Fig.~\ref{abundfig} the abundances in the two components of the
IRAS2A outflow, in the IRAS2A envelope and L1157 outflow are
compared. In particular SiO, CH$_{3}$OH, SO and CS are significantly
enhanced by factors 10-10$^4$ in the outflow regions compared to the
envelope and quiescent dark cloud. The abundances in the
\object{IRAS2A} wing component and \object{L1157} agree very well for
SO and SiO, but the abundances of HCN, H$_2$CO and CS are lower in the
\object{IRAS2A} outflow by factors of 10-100. The CH$_{3}$OH abundances
are very large for both outflows - between 5 and 10\% of that of
CO. This is especially significant in comparison with abundances found
in the Orion hot core and the low-mass protostar hot core in
IRAS~16293-2422. There CH$_{3}$OH is thought to be enhanced through
thermal evaporation off dust grain ice mantles, but its abundance is a
factor of 10-100 lower than in the outflow regions. Also the SiO
abundances are different between the hot core sources and the outflow
regions, whereas the SO abundances are practically identical. For CS
and HCN, the abundances have to be significantly higher in the inner
regions of IRAS~16293-2422 in order to match the outflow
abundances. Compared to the dark cloud, L134N, the CH$_{3}$OH abundances
again stand out as significantly increased, together with CS and
SO. HCN and H$_2$CO have similar abundances in the outflow region and
the dark cloud whereas HCO$^+$ shows slightly lower abundances in the
outflow regions.
\begin{table*}
\caption{Abundances for the two components of the \object{IRAS2A}
outflow compared to other outflows and protostellar
environments.}\label{abundtable}
\begin{tabular}{lll|llllll}\hline\hline
Molecule        & \object{IRAS2A} core$^a$  & \object{IRAS2A} wing$^a$ & \object{L1157}$^b$ & HH2$^c$ & BHR71$^d$ & IRAS2A env.$^e$ & IR16293$^{f}$ & L134N$^g$ \\ \hline
CO              & =1$\times 10^{-4}$     & =1$\times 10^{-4}$  & =1$\times 10^{-4}$ & =1$\times 10^{-4}$ & =1$\times 10^{-4}$ & 2$\times 10^{-5}$    & 4$\times 10^{-5}$          & =1$\times 10^{-4}$ \\
CH$_3$OH        & 6.7$\times 10^{-7}$    & 4.8$\times 10^{-6}$ & 2$\times 10^{-5}$  & 2$\times 10^{-8}$  & 2$\times 10^{-7}$  & 2$\times 10^{-9}$    & 3$\times 10^{-7}$$\dagger$ & 8$\times 10^{-9}$  \\ 
CS              & 1.1$\times 10^{-8}$    & 2.7$\times 10^{-8}$ & 2$\times 10^{-7}$  & 7$\times 10^{-10}$ & 6$\times 10^{-9}$  & 3$\times 10^{-9}$    & 3$\times 10^{-9}$          & 1$\times 10^{-9}$  \\
HCN             & 2.2$\times 10^{-9}$    & 4.8$\times 10^{-9}$ & 5$\times 10^{-7}$  & 1$\times 10^{-9}$  & -          & 2$\times 10^{-9}$    & 1$\times 10^{-9}$          & 7$\times 10^{-9}$  \\
HCO$^+$         & 1.3$\times 10^{-9}$    & 2.9$\times 10^{-9}$ & 3$\times 10^{-8}$  & 3$\times 10^{-8}$  & 9$\times 10^{-10}$ & 3$\times 10^{-9}$    & 1$\times 10^{-9}$          & 8$\times 10^{-9}$  \\
H$_2$CO         & 1.7$\times 10^{-8}$    & 1.5$\times 10^{-8}$ & 3$\times 10^{-7}$  & 2$\times 10^{-8}$  & -          & 8$\times 10^{-10}$   & 6$\times 10^{-8}$$\dagger$ & 2$\times 10^{-8}$ \\
N$_2$H$^+$      & 1.8$\times 10^{-9}$    & $\ldots$    & $\ldots$   & -          & -          & 5$\times 10^{-9}$    & 1$\times 10^{-10}$         & 6$\times 10^{-10}$ \\
SiO             & $\ldots$       & 1.1$\times 10^{-7}$ & 7$\times 10^{-8}$  & $\ldots$   & 7$\times 10^{-10}$ & $<$5$\times 10^{-11}$& 5$\times 10^{-9}$$\dagger$ & $<$1$\times 10^{-11}$$^{h}$ \\
SO              & 1.2$\times 10^{-8}$    & 1.2$\times 10^{-7}$ & 3$\times 10^{-7}$  & 8$\times 10^{-9}$  & -          & 3$\times 10^{-9}$    & 3$\times 10^{-7}$$\dagger$ & 6$\times 10^{-9}$  \\ \hline
\end{tabular}

Notes: ``-'' Molecule not observed. ``$\ldots$'' Molecule observed but not detected.\\
$^a$Core and wing part of IRAS2A outflow (this paper). \\ 
$^b$The L1157 outflow \citep{bachiller97} \\
$^c$The HH2 molecular condensation \citep{girart02} \\
$^d$The BHR71 outflow \citep{garay98} \\
$^e$The IRAS2A protostellar envelope \citep{jorgensen02,paperii}\\
$^f$The IRAS~16293-2422 envelope \citep{schoeier02}. ``$\dagger$''
indicate abundances for the warm inner part of the envelope; other
abundances are averages over the entire envelope.\\
$^g$The ``C'' position of the L134N dark cloud \citep{dickens00} \\ 
$^{h}$Derived using upper limits on [SiO]/[HCN] abundance ratio for L134N of 0.0015 from
\cite{ziurys89}.
\end{table*}

\subsection{Dynamical time scales}
The dynamical time-scale of the east-west outflow can be estimated by
assuming that: (1) the terminal velocity of the outflow is equal to
the maximal radial velocity of the SiO emission, i.e., 29~km~s$^{-1}$,
(2) the rest velocity of the cloud corresponds to the velocity of the
narrow non-shocked features ($V_{\rm LSR} = 7$~km~s$^{-1}$), (3) the
distance between the tip of the SiO outflow and the continuum position
is the full extent of the outflow and (4) the inclination between the
plane of the sky and the outflow is small, as its high degree of
collimation and extent plus the relatively low observed velocities
seem to indicate. With these assumptions one finds a dynamical age:
\[t_{\rm dyn}\approx 4\times 10^3 \times \left(\frac{d}{220\, {\rm pc}}\right) \,{\rm years,}\]
which is in good agreement with the dynamical timescale of 
\[t_{\rm dyn}\approx (3-7) \times 10^3 \times \left(\frac{d}{220\, {\rm pc}}\right) \,{\rm years} \]
found by \cite{bachiller98} from their CH$_3$OH maps.

The dynamical timescale calculated this way is subject to significant
systematic errors. The true extent of the outflow will be $L_{\rm
true}=L_{\rm obs}\cos i$, while the terminal velocity of the outflow
relates to the observed maximum radial velocity as $V_{\rm
true}=V_{\rm obs}\sin i$, modifying the dynamical timescale by a
factor $\tan i$. Furthermore the dynamical timescale at best reflects
the properties of the outflow at the present moment - changes in the
flow velocities throughout the history will similarly change the
outflow dynamical timescale. Thus, it is definitely not an unbiased
indicator of the age of the driving protostellar source. Still, it
agrees well with the timescales derived from comparison between the
envelope structure and collapse models \citep{n1333i2art} and
therefore does give an indication of the order of magnitude of the
appropriate timescale to be used when discussing the chemical
evolution of the shock in the following section.

\subsection{Chemical evolution}
The above analysis shows that a number of species are significantly
enhanced in the outflow region. The various spectral signatures and
morphologies in the interferometer maps together with the varying
abundances found in the \object{L1157} and \object{IRAS2A} outflows
indicate different mechanisms regulating the abundances. For the
sulfur-bearing species, i.e., CS and SO, the enhancement is thought to
occur as a result of enhanced H$_2$S formation as seen in hot cores
\citep{pineaudesforets93,charnley97}. CS and SO show different
morphologies in the interferometry maps and different spectral
signatures. In particular, CS is present in a narrow high velocity
component, while SO only shows up at lower velocities and with more
compact emission. In this context the differences and similarities
between the abundances found in the \object{L1157} and \object{IRAS2A}
outflows are also interesting (see Fig.~\ref{abundfig}): SiO and SO
are seen to be quite similar in the \object{L1157} outflow and
\object{IRAS2A} wing component, while CS, H$_2$CO and HCN are found to
be 10 to 100 times more abundant in the \object{L1157} outflow. An
explanation could be differences in the abundances of atomic carbon:
the production of CS is increased with a higher abundance of atomic
carbon, whereas the SO is more closely related to the oxygen abundance
\citep[e.g.,][]{vandishoeck98}. In models for gas-phase chemistry, CS,
HCN and H$_2$CO are all enhanced with higher abundances of atomic
carbon, thus potentially explaining the difference between the
\object{IRAS2A} and \object{L1157} outflows. The atomic carbon could
either be produced in the shock itself or by photodissociation of CO
in the pre-shocked gas \citep[e.g., as in the case of
IC~443,][]{keene96}.

\begin{figure}
\resizebox{\hsize}{!}{\includegraphics{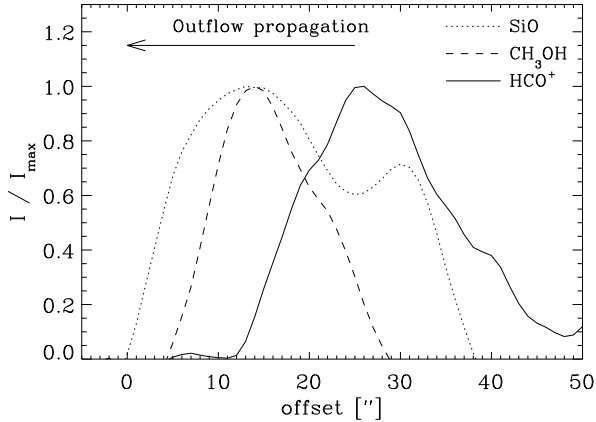}}
\caption{Spatial differences between SiO, CH$_3$OH and HCO$^+$ along
the shock propagation from the interferometer maps. The emission from
each species has been integrated in strips perpendicular to an axis
aligned in the propagation direction of the outflow (position angle of
19$^\circ$ with the RA axis and with zero-point at (87'',-23'') as in
Fig.~\ref{pv_diagram}). The SiO and CH$_3$OH data cubes have been
reduced to the same spatial resolution as that of the HCO$^+$
data.}\label{shock_propagation}
\end{figure}

The enhancement of SiO is thought to be caused by atomic silicon
sputtering from the surfaces of dust grains and quickly forming SiO in
the gas-phase through reactions with OH
\citep[e.g.,][]{schilke97,caselli97,pineaudesforets97}. CH$_{3}$OH, H$_2$CO
and HCN, in contrast, are most likely enhanced through direct
evaporation of ice mantles. Alternative explanations, e.g., gas-phase
reactions between CH$_3^+$ and H$_2$O forming CH$_3$OH only produce
CH$_3$OH abundances of $1-5\times 10^{-8}$ \citep[e.g.,][]{millar91},
significantly lower than those found in the outflow regions of
$10^{-6}-10^{-5}$ \citep[this
paper,][]{bachiller95,bachiller97,bachiller98}. The spectral
signatures of CH$_3$OH compared to SiO indicate that it is formed at
lower velocities. Furthermore, it is seen that CH$_3$OH peaks slightly
further downstream compared to, e.g., SiO in the interferometer
maps. This is clearly illustrated in Fig.~\ref{shock_propagation}
where the intensities of SiO, CH$_3$OH and HCO$^+$ are compared along
the outflow propagation axis. \cite{garay00} analyzed regions of
different shock velocities in the outflow associated with NGC~2071 and
found that CH$_{3}$OH was most prominent in regions with low shock
velocities ($v_s \lesssim 10$~km~s$^{-1}$). \citeauthor{garay00} suggested
that since molecules such as CH$_{3}$OH, H$_2$CO and HCN are more
volatile, they would only be capable of surviving at much lower shock
velocities than would, e.g., SiO. Applied to the \object{IRAS2A} outflow
this explains both the differences in terminal velocities between SiO
on the one hand and HCN, H$_2$CO and CH$_{3}$OH on the other, but also
the different ``onsets'' along the propagation direction between
CH$_{3}$OH and SiO from the interferometer maps as seen in
Fig.~\ref{shock_propagation}.

The characteristic core-wing structure of the observed lines is
similar to that seen in a SiO survey of protostellar outflows by
\cite{codella99}, who argued that it could be an evolutionary effect
with the SiO being produced at high velocities and subsequently slowed
down toward lower velocities. They argued that it would take $\sim
10^4$~years to slow down an outflow-induced shock, which is similar to
the time it would take SiO to be destroyed either through direct
accretion onto dust grains \citep{bergin98} or through reactions with
OH, forming SiO$_2$ \citep{pineaudesforets97}. If this picture applies
to the \object{IRAS2A} outflow, it is not surprising that SiO has low
abundances in the ``core component'' - or ambient cloud. Since the SiO
destruction timescale is similar to the dissipation timescale for the
protostellar shock, SiO is almost completely destroyed in the
slow-down phase and will therefore not be seen in the
low-velocity/quiescent component. On the other hand, since SiO is
created as a direct result of the shock impact
\citep{pineaudesforets97,schilke97}, it traces the highest velocities
in the outflow, together with CO.  The characteristic molecular
depletion timescale at densities of $10^6$~cm$^{-3}$ typical of the
outflow region (Sect.~\ref{stateq}) is on the order of 10$^3$~years,
comparable to the outflow dynamical timescale, and could therefore
explain why, e.g., SiO and CH$_3$OH are not observed over the entire
extent of the outflow back to the central protostar.

The differences between the hot core/warm envelope and outflow
abundances of, e.g., CH$_{3}$OH and SiO could be caused by differing time
scales related to the densities in the differing regions: the density
in the hot inner part of protostellar envelopes is higher by 2-3
orders of magnitude than what is found in the outflow regions. This
will lead to more rapid destruction of molecules with ``anomalous''
abundances, e.g., SiO, either through accretion or reactions with other
species and therefore also lower abundances in the envelope
regions. Of course the mechanisms for producing the given molecules in
the first place are also likely to be dependent on the environment,
further complicating the picture.

The depletion timescale for CH$_3$OH may also be taken as an important
clock related to the HCO$^+$ abundance. As noted previously, HCO$^+$
stands out compared to the other molecules tracing material only in
the aftermath of the shock. In the \object{L1157} outflow, HCO$^+$ was
only found to be prominent in the part of the outflow close to the
driving source. Through chemical models, \cite{bergin98} found that
HCO$^+$ should be destroyed after the passage of the shock through
reactions with H$_2$O (${\rm H}_2{\rm O}\,+\,{\rm HCO}^+\,
\rightarrow\, {\rm H}_3{\rm O}^+\, +\, {\rm CO}$), but increases later
as the water abundance reaches lower levels due to freeze-out. This is
in fact seen in interferometer data as illustrated in
Fig.~\ref{shock_propagation}: the emission of HCO$^+$ and CH$_3$OH is
almost anticorrelated, with CH$_3$OH being located closer to the
``head'' of the outflow and HCO$^+$ showing up in the aftermath of the
shock. As higher abundances of both CH$_3$OH and H$_2$O are expected
to be results of grain mantle evaporation and the timescales for their
freeze-out are similar, the HCO$^+$ and CH$_3$OH enhancements should
indeed be anticorrelated as seen in Fig.~\ref{shock_propagation}.

N$_2$H$^+$, like HCO$^+$, is expected to be destroyed by reactions with
H$_2$O. In contrast to HCO$^+$, however, CO may also be important in
destroying N$_2$H$^+$ \citep[e.g.,][]{bergin97,charnley97}. Observational
studies of pre- and protostellar objects
\citep[e.g.,][]{bergin01,tafalla02,n1333i2art} suggest that N$_2$H$^+$ is
enhanced where CO is depleted. The narrowness of the N$_2$H$^+$ lines and
the morphology of the emission in the \object{IRAS2A} region indicate
that this molecule is indeed only tracing the ambient cloud material
where CO may be depleted and not the outflowing material where CO is
returned to the gas-phase.

\section{Conclusion\label{conclusion}}
A (sub)millimeter study of the shock associated with the
\object{NGC1333-IRAS2A} outflow has been presented. Both single-dish
and interferometry line observations are presented, which allows for a
detailed discussion of both the physical and chemical properties in
the shocked region and the spatial distribution of emitting
species. The main findings are as follows:

\begin{enumerate}
\item Interferometer observations of the outflow region reveal a
distinct morphology with a narrow high velocity feature in CS, SiO and
HCN, while the low velocity part traces the more spatially extended
material. HCO$^+$ is seen only in the aftermath of the shock, whereas
N$_2$H$^+$ shows very narrow lines and does not seem to be present in
the outflowing gas.

\item Statistical equilibrium calculations show that the region can be
divided into two distinct components: low velocity material in the
ambient, quiescent cloud and high velocity material associated with
the outflow. It is found that the conditions in the quiescent material
are consistent with a temperature of $\sim$20~K and density of
$1\times 10^6\, {\rm cm}^{-3}$, while the shocked gas is warmer with a
temperature of $\sim$70~K and density of $2\times 10^6\, {\rm
cm}^{-3}$. Within these components, however, the physical conditions
are remarkably homogeneous as indicated by the similarities between
the lineshapes for different transitions of various molecules and
between the wing emission in spectra from interferometry and
single-dish observations.

\item The chemistry in the outflow and quiescent regions are
significantly different. CH$_3$OH, SiO and the sulfur-bearing species
are significantly enhanced in the outflowing gas, whereas HCN,
H$_2$CO, HCO$^+$ and N$_2$H$^+$ show abundances more similar to those
found in molecular clouds and protostellar envelopes. Compared to the
well-studied \object{L1157} outflow the CS, HCN and H$_2$CO abundances
are markedly lower. This could be due to differences in the amount of
atomic carbon in the shocked or pre-shocked gas. Higher abundances of
the shock-tracing molecules (in particular SiO and CH$_{3}$OH) in outflows
compared to the inner warm envelopes may be related to more rapid
destruction of these molecules in the envelopes where the densities
are higher.

\item A scenario is suggested where the highly collimated protostellar
outflow is progressing into a region with a steep density gradient,
focusing the shock and giving rise to the narrow morphology
observed. This leads to a shock-induced chemistry, which can explain
both the morphologies and qualitatively the abundances of the
different molecules. In particular, CH$_{3}$OH is seen to be greatly
enhanced reaching an abundance of about 5\% of the observed CO
abundance in the shocked gas. CH$_{3}$OH is, however, not observed to as
high velocities as seen in SiO, possibly as a consequence of CH$_{3}$OH
being more volatile. Thus together with, e.g., HCN and H$_2$CO,
CH$_{3}$OH is not able to survive at higher velocities.
\end{enumerate}

This work illustrates the large impact of protostellar outflows in
shaping the physical and chemical properties of their parental
environment. The combination of high-resolution interferometer
observations and single-dish spectra makes it possible to address the
physical and chemical conditions in the shocked and ambient gas and to
investigate the spatial variation and time-scales characteristic for
the shock induced chemistry. So-far only a few shocks have been
studied in great chemical detail. Similar systematic studies of a
large number of different outflows will allow for a more detailed
comparison between outflows and shocks of different velocities and
energetics and in different environments. Future observations with
facilities such as the SMA, CARMA, and ALMA will allow further studies
of the variation of physical and chemical conditions in shocks through
high resolution, high sensitivity multi-transition molecular line
observations. Also high spatial resolution observations of H$_2$O
lines with Herschel-HIFI can confirm the anticorrelation between
HCO$^+$ and H$_2$O. All such more detailed observational studies will
serve as important starting points for more detailed physical and
chemical models for shocks in protostellar environments.

\begin{acknowledgements}
We thank the referee for a prompt and well-considered report. The
research of JKJ is funded by the Netherlands Research School for
Astronomy (NOVA) through a network 2 Ph.D. stipend and research in
astrochemistry in Leiden is supported by a Spinoza grant. GAB
acknowledges support from the NASA Origins of Solar Systems
program. FLS further acknowledges financial support from the Swedish
Research Council.
\end{acknowledgements}

\bibliographystyle{aa}
\end{document}